\documentclass[twocolumn,reprint,aps,prd]{revtex4-1}
\usepackage[utf8]{inputenc}
\usepackage[english]{babel}
\usepackage[toc,page]{appendix}
\usepackage{amsmath}
\usepackage{mathrsfs}
\usepackage{amsfonts}
\usepackage{amssymb}
\usepackage{amsthm}
\usepackage{bbm}
\usepackage{hyperref}
\usepackage{graphicx}

\begin{document}
\title{Schwinger pair production and string breaking in non-Abelian gauge theory from real-time lattice improved Hamiltonians}
\author{Daniel Spitz, Jürgen Berges}
\affiliation{Institut für theoretische Physik, Universität Heidelberg, Philosophenweg 16, 69120 Heidelberg, Germany}
\begin{abstract}
Far-from-equilibrium dynamics of $SU(2)$ gauge theory with Wilson fermions is studied in \mbox{$1+1$} space-time dimensions using a real-time lattice approach. Lattice improved Hamiltonians are shown to be very efficient in simulating Schwinger pair creation and emergent phenomena such as plasma oscillations. As a consequence, significantly smaller lattices can be employed to approach continuum physics in the infinite-volume limit as compared to unimproved implementations. This allows us to compute also higher-order correlation functions including four fermion fields, which give unprecedented insights into the real-time dynamics of the fragmentation process of strings between fermions and antifermions.
\end{abstract}
\maketitle

\section{Introduction}
Confinement in quantum chromodynamics (QCD) manifests itself, amongst others, in that the energy stored in a gluon string between a quark and an antiquark rises linearly with the string length. A critical distance between the quark and the antiquark exists, at which the string starts to break through fermion-antifermion pair production, reducing the energy stored in the string~\cite{Philipsen:1998de, Knechtli:1998gf, Bali:2005fu, Pepe:2009in}. The real-time dynamics of the nonperturbative string formation and fragmentation process, however, poses serious obstacles to a computation from first principles. In general, the nonequilibrium dynamics of quantum fields is not amenable to a formulation in Euclidean space-time and the use of importance sampling techniques, such that alternative approaches have to be employed.

Already the dynamics of geometric confinement and string breaking for gauge theories in $1+1$ space-time dimensions is very rich with a hierarchy of timescales and a close link to the Schwinger pair production mechanism, sharing important aspects with their higher dimensional counterparts. In $1+1$ dimensions, Abelian string dynamics and the phenomenon of multiple string breaking from supercritical field strength has been studied using classical-statistical reweighting techniques on the lattice~\cite{Hebenstreit:2013baa}. For gauge theories with fermions these methods have also been employed to fermion production in $1+1$~\cite{Hebenstreit:2013qxa} and $3+1$ space-time dimensions~\cite{Kasper:2014uaa} using staggered fermions, as well as Wilson fermions~\cite{Saffin:2011kc,Saffin:2011kn,Gelfand:2016prm,Tanji:2017xiw,Zache:2018jbt}. For real-time Wilson fermions, lattice improvement techniques have been shown to be extremely useful in studies of quantum anomalies in QCD~\cite{Mueller:2016ven,Mace:2016shq} and QED~\cite{Mueller:2016aao} in $3+1$ dimensions.

Much of the recent interest in the low-dimensional dynamics comes from the prospect of studying important aspects of the gauge theory using quantum simulators~\cite{Szpak:2011jj, Banerjee:2012pg, Banerjee:2012xg, Tagliacozzo:2012vg, Zohar:2012xf, Zohar:2012ay, Zohar:2011cw}, with a first proof-of-principle implementation on a trapped-ion quantum computer~\cite{Martinez:2016yna}. In this context, the dynamics of Abelian gauge theory has been studied in great detail using classical statistical reweighting techniques~\cite{Kasper:2015cca,Kasper:2016mzj,Zache:2018jbt}. Important developments concerning string-breaking dynamics in Abelian and non-Abelian gauge theory have also been based on tensor network techniques in $1+1$ dimensions~\cite{Buyens:2015tea, Kuhn:2015zqa, Pichler:2015yqa, Buyens:2016hhu} and recently on a Gaussian variational ansatz~\cite{Sala:2018dui}. Additionally, the dynamics of flux strings has been studied in the $U(1)$ gauge-Higgs model~\cite{Kuno:2016ipi}. Compared to the Schwinger model, $(1+1)$-dimensional $SU(N)$-theory with massive fermions shares with full QCD the existence of confining strings between color charges as well as mesonic bound states in its spectrum~\cite{ABDALLA1996253, GROSS1996109, Abdalla1997, Frishman:1997uu}. Non-Abelian Schwinger pair production exhibits several distinct features such as interference and cancellations of currents between the different participating color charges~\cite{Tanji:2010eu, Tanji:2015ata, Tanji:2016dka}.

In the present work we investigate for the first time the process of non-Abelian Schwinger pair production and the dynamics of string breaking in $1+1$ dimensions using classical-statistical reweighting techniques. For Wilson fermions, we analyze the use of lattice improved Hamiltonians in a Kogut-Susskind-like real-time approach. We demonstrate that next-to-leading order (NLO) and next-to-next-to-leading order (NNLO) improvements in the fermion sector of the Hamiltonian can be efficiently employed to significantly reduce the lattice sizes required to approach the continuum and infinite-volume limit of the results. This seems crucial in view of the prospect of implementing these theories on a quantum simulator with a limited number of qubits. Moreover, this allows us to compute higher correlation functions involving four fermion fields. Together with the chromoelectric field strength, color charge and fermion number distributions, these charge-charge correlations provide a very detailed picture of the string dynamics and fragmentation process far from equilibrium.

This publication is structured as follows: In Sec.~\ref{SectionRealTimeLatticeSetup} the real-time lattice setup is developed, including field equations of motion and formulas for the computation of various observables from the lattice gauge theory. Subsequently, in Sec.~\ref{SectionHomogeneous} we focus on coherent chromoelectric fields that give rise to a homogeneous production of fermions. The available analytics provide a stringent precision test for the simulated lattice improved fermion number momentum spectra and fermion production rates. We explore thoroughly the occurring non-Abelian plasma oscillations, including an analysis of the effects of lattice improvements up to second order. In Sec.~\ref{SectionInhomogeneous} the breaking of color strings is investigated. Beginning with a description of corresponding initial conditions, we move on to a phenomenological analysis of the processes that take place in breaking color strings. Observing various dynamical color charge accumulations between the static external charges, we study charge-charge correlations among them and the propagation characteristics of such correlations. In Sec. \ref{SectionConclusions} we summarize, draw conclusions and give an outlook.

\section{Real-time lattice gauge theory with Wilson fermions}\label{SectionRealTimeLatticeSetup}

The Lagrangian density of $SU(2)$ gauge theory with gauge potential $A_\mu(t,x)$ and one fermion flavor $\psi(t,x)$ transforming in the fundamental representation of the gauge group reads in continuous space-time:
\begin{equation}
\mathcal{L} = -\frac{1}{2}\,\text{tr}\,F^2 + \overline{\psi}\big(i\gamma^\mu \partial_\mu -g \gamma^\mu A_\mu- m \big)\psi\,.
\end{equation}
In $(1+1)$ space-time dimensions there are no color-magnetic fields, and the field-strength tensor $F_{\mu\nu}$ is determined by the color-electric field $E(t,x)$ with nontrivial components  
\begin{equation}
F_{10}(t,x) = -F_{01}(t,x) = E(t,x) = E^a(t,x)T^a,
\end{equation}
where the summation over gauge indices $a=1,2,3$ is implied and $T^a$ denote the gauge group generators.
In what follows temporal-axial gauge is employed, i.e.\ $A_0(t,x)=0$, thus denoting $A(t,x)\equiv A_1(t,x)$. Since gluon self-couplings vanish in temporal-axial gauge in one spatial dimension, the gauge dynamics simplifies tremendously. 

The $(1+1)$-dimensional Dirac algebra is composed of two Dirac gamma matrices,
\begin{equation}
\{\gamma^\mu,\gamma^\nu\} = 2\eta^{\mu\nu},\,\,\,\,\text{with}\,\,\,\,(\gamma^\mu)^\dagger = \gamma^0\gamma^\mu\gamma^0,
\end{equation}
with Minkowski metric $\eta^{\mu\nu}=\text{diag}(1,-1)$. Pauli matrices serve as a representation of this algebra: $\gamma^0\equiv \sigma_1$ and $\gamma^1 \equiv -i \sigma_2$. The chirality matrix then reads
\begin{equation}
\{\gamma^\mu,\gamma_5\} = 0,\,\,\,\,\text{with}\,\,\,\,(\gamma_5)^\dagger = \gamma_5,\, (\gamma_5)^2 = 1,
\end{equation}
which is defined using the third Pauli matrix $\gamma_5\equiv \sigma_3$.

\subsection{Lattice implementation with improved Hamiltonian}
To formulate the theory on the lattice we discretize the spatial degree of freedom,
\begin{equation}
\Lambda = \left\{n=x/a\in \{0,\dots, N_1 -1\} \right\},
\end{equation}
such that positions may be labeled by integers $n=x/a$. The total number of lattice sites is $N_1$, the lattice spacing $a$. The lattice has total length $L=N_1a$. We employ spatially periodic boundary conditions. We neither discretize the time direction nor impose temporal periodicity assumptions. In momentum space we find the inverse lattice
\begin{equation}
\tilde{\Lambda} = \left\{\tilde{q}\equiv q-N_1/2=\frac{Lp}{2\pi}\in \{-N_1/2,\dots, N_1/2-1\}\right\}.
\end{equation}
Spatial link variables are constructed as
\begin{equation}
U_n(t) \equiv \exp \big(igaA_n(t)\big),
\end{equation}
parallel transporting chromoelectric flux from site $n$ to site $n+1$. Temporal link variables are identically unity in temporal-axial gauge. Inverse link variables are given by $U^\dagger_{n+1} (t)$, parallel transporting chromoelectric flux from site $n+1$ to site $n$. For fermion fields and link variables the following behavior under a given gauge transformation $V_n(t)\in SU(2)$ is observed:
\begin{subequations}
\begin{align}
\psi^{\mathstrut}_n(t) &\mapsto  V^{\mathstrut}_n(t)\, \psi^{\mathstrut}_n(t)\,,\\
U^{\mathstrut}_n(t) &\mapsto  V_n^{\mathstrut}(t) \,U^{\mathstrut}_n (t)\,V^\dagger_{n+1} (t)\,.
\end{align}
\end{subequations}
For later use we construct products of neighboring link variables with integer $z>0$,
\begin{subequations}
\begin{align}
U_{n,z}(t) &\equiv  U_n(t)\,U_{n+1}(t)\cdots U_{n+z-2}(t)\,U_{n+z-1}(t)\,,\\
U_{n,-z}(t) &\equiv  U^\dagger_{n-1}(t)\,U^\dagger_{n-2}(t)\cdots U^\dagger_{n-z+1}(t)\,U^\dagger_{n-z}(t)\,.
\end{align}
\end{subequations}
Nontrivial path-ordering of gauge group elements applies here.
The gauge sector Hamiltonian of our model reads
\begin{equation}\label{HamiltonianBosons}
H^{(\text{g})}(t) = a \sum_{n\in \Lambda} \text{tr}\,E_n^2(t) = \frac{a} {2} \sum_{n} E^a_n E^a_n\,.
\end{equation}

To implement spinor fields on the lattice without doubler excitations we employ Wilson fermions~\cite{Saffin:2011kn, Saffin:2011kc, Rothe2012_LatticeGaugeTheories}. Working in the Hamiltonian formulation of lattice gauge theory, the fermion sector Hamiltonian is given by
\begin{align}\label{HamiltonianFermions}
H^{(\text{f})}(t) =\,\, &  a\sum_{n}\psi_{n}(t)^{\dagger}\gamma^{0}\bigg[ (m+Kr/a)\,\psi_{n}(t)\nonumber\\
& -\frac{1}{2a}\sum_{k=1}^{K}C_k\left(i\gamma^{1}+kr\right)U_{n,k}(t)\,\psi_{n+k}(t) \nonumber\\
& + \frac{1}{2a}\sum_{k=1}^{K}C_k\left(i\gamma^{1}-kr\right)U_{n,-k}(t)\,\psi_{n-k}(t)\bigg].
\end{align}
Here $r$ is the Wilson parameter with $0 < r$, and the integer $K$ denotes the order of the lattice improvement for the Hamiltonian: Using an appropriate choice of coefficients $C_k$ for $k = 1, \ldots K$ it is possible to cancel certain lattice artifacts of order $\mathcal{O}(a^{2K-1})$ in the fermion lattice Hamiltonian \cite{Mace:2016shq}. Choosing $C_1=1$ and all other coefficients to vanish, one recovers the unimproved Wilson Hamiltonian, which is accurate to $\mathcal{O}(a)$ and which we call leading order (LO). Using $C_1=4/3$ and $C_2=-1/6$ we obtain $\mathcal{O}(a^3)$-accuracy, labeled NLO. Including a third nonvanishing term, $C_1=3/2,\,C_2=-3/10,\,C_3=1/30$, we obtain $\mathcal{O}(a^5)$-accuracy, labeled NNLO.

The complete Hamiltonian including gauge and fermion fields reads
\begin{equation}\label{FullHamiltonian}
H(t)=H^{(\text{g})}(t)+H^{(\text{f})}(t).
\end{equation}

\subsection{Initial conditions and classical-statistical reweighting}
To solve the Cauchy problem of the time evolution of field degrees of freedom, we need to specify both fermion and gauge initial conditions at time $t=0$. We assume that initially the two sectors decouple, subsequently quenching the system into a coupled state via time evolution with the full Hamiltonian (\ref{FullHamiltonian}). For the physics of fermion production from strong gauge fields or color charges, fermions are initialized as free fermions throughout this work. Details on the fermion initial conditions are given in Appendix \ref{AppendixFermionicInitialConditions}.

While the gauge field initial conditions are specified in more detail in Secs. \ref{SectionHomogeneous} and \ref{SectionInhomogeneous}, we consider strong fields for which the initial color-electric fields are of order of the critical field strength $E_c = m^2/g$. In this case, well-established classical-statistical reweighting techniques can be employed to replace the full quantum dynamics to very good accuracy by sampling classical gauge field dynamics~\cite{Aarts:1998td, Aarts:2000mg, Aarts:2001yn, Berges:2007ym, Berges:2010zv}. Observables are then computed as ensemble averages of the results from a sufficiently large number of sampling runs until convergence is observed. For details on the sampling of gauge field quantum initial conditions we refer to Appendix \ref{AppendixBosonicQuantumFluctuations}.

To this end, gauge degrees of freedom, i.e.\ $U_n$ and $E_n$, are evolved classically from given gauge field configurations and fermion correlations. To resolve the manifest quantum nature of the fermions, the fermion equation of motion is solved on operator level, applying a mode-function expansion:
\begin{equation}
\psi_n (t) = \frac{1}{L}\sum_{q\in \tilde{\Lambda}}\sum_{a^{\mathstrut}=1}^2 \big[\phi_{n,q,a}^u(t) \,b_{q,a}^{\mathstrut} + \phi_{n,q,a}^v(t)\, d^\dagger_{q,a}\big],
\end{equation}
with time-dependent mode functions $\phi_{n,q,a}^u(t),\phi_{n,q,a}^v(t)$. The time-independent creation and annihilation operators $b_{q,a}$ and $d_{q,a}^\dagger$, respectively, satisfy
\begin{equation}
\big\{b_{q,a}^{\mathstrut},b_{p,b}^\dagger\big\} = \big\{ d_{q,a}^{\mathstrut},d_{p,b}^\dagger\big\} = L\,\delta_{q,p}\,\delta_{a,b}\,.
\end{equation}
Fermion occupation numbers are determined by
\begin{subequations}
\begin{align}
\big\langle b^\dagger_{q,a} b_{q,a}\big\rangle &=\, L\, n^u_{q,a}\,,\\
\big\langle d^\dagger_{q,a} d_{q,a}\big\rangle &=\, L\, n^v_{q,a}\,.
\end{align}
\end{subequations}
We construct the statistical propagator on the lattice,
\begin{equation}\label{eq:statprop}
\Delta_{m,n}(t) = \frac{1}{2}\, \langle [\psi_m (t), \,\overline{\psi}_n(t)] \rangle \,,
\end{equation}
which reads in terms of mode functions,
\begin{eqnarray}
\Delta_{m,n}^{ab;\alpha\beta}(t) &=& \frac{1}{2L} \sum_{q\in \tilde{\Lambda}} \bigg(\phi_{m,q,a}^{u, \alpha} (t)\,\overline{\phi}_{n,q,b}^{u,\beta} (t)\left(1-2\,n^{u}_{q,a}\right)\nonumber \\
&& - \phi_{m,q,a}^{v, \alpha} (t)\, \overline{\phi}_{n,q,b}^{v,\beta} (t) \left(1-2\,n^{v}_{q,a}\right)\bigg).\label{StatisticalPropagator}
\end{eqnarray}
The temporal derivative of an operator $O$ may be calculated using
\begin{equation}
i\partial_t O = [O,H]\,.
\end{equation}
Key steps in the derivation of each of the following time-evolution equations are given in Appendix \ref{AppendixEOMS}. For the chromoelectric field one obtains the equation
\begin{widetext}
\begin{equation}\label{EOMChromoelectricField}
\partial_t E_n^a(t) = g\sum_{k=1}^{K} \sum_{m=1}^kC_k \,\text{Re}\, \text{Tr} \big(\Delta_{n+m,n+m-k}(t)\,(\gamma^1-ikr)\, U_{n+m-k,k-m}(t)\,T^a\, U_{n,m}(t)\big)\,,
\end{equation}
\end{widetext}
where the trace runs over color and Dirac indices.
The time evolution of the link variable follows from
\begin{equation}\label{EOMLinkVariables}
\partial_t U_n = iga E_n U_n\,,
\end{equation}
with $E_n$ and $U_n$ acting upon each other via the standard matrix product.
Finally, the fermion field operators obey
\begin{align}\label{EOMFermionModes}
\partial_t \psi_n =\,\,& -i\gamma^0 \left(m+K\frac{r}{a}\right)\,\psi_n \nonumber\\
& - \frac{1}{2a}\sum_{k=1}^{K}C_k\gamma^0(\gamma^1 - ikr) \, U_{n,k}(t)\,\psi_{n+k}(t) \nonumber\\
& + \frac{1}{2a}\sum_{k=1}^{K}C_k\gamma^0(\gamma^1 + ikr) \,U_{n,-k}(t)\,\psi_{n-k}(t)\,.
\end{align}
By linear independency of the creation and annihilation operators $b_{q,a}, d^\dagger_{q,a}$ the equation of motion for each and every mode function takes this form. To solve the given system of equations, we specify a time-step width $a_t$ and employ a four-step Runge-Kutta algorithm.

\subsection{Abelianization for homogeneous fields}\label{SubsectionAbelianization}
Diagonalizing gauge degrees of freedom in color space, which is generally possible by a local gauge transformation, provides a way to simplify the $SU(2)$ gauge group structure. It results in a $U(1)\times U(1)$ gauge theory, which may be understood intuitively better compared to the $SU(2)$ theory~\cite{Tanji:2010eu, Tanji:2015ata}. 
For later interpretation of lattice results in the context of Schwinger pair production, we consider in the following the Abelianization procedure for homogeneous field configurations. 

Let $E_n(t)$ be homogeneous with no rotations in color space taking place. We write for all $n\in\Lambda$,
\begin{equation}
E_n(t)=E^a(t) T^a = E(t) \, n^a T^a,
\end{equation}
with a constant vector $n^a$, such that $n^a n^a = 1$. By Hermiticity of $n^a T^a$ there exists a unitary matrix $U$, such that
\begin{equation}\label{DiagonalizedcolorDirectionMatrix}
U n^a T^a U^\dagger = \text{diag} \big(1/2,-1/2\big)\,.
\end{equation}
Explicitly, $U$ may be given by
\begin{equation}\label{DiagonalizationMatrixU}
U = \frac{1}{\sqrt{2}}\left(\begin{matrix}
+(n^1+in^2)/\sqrt{1-n^3} &  \sqrt{1-n^3}\\
-(n^1+in^2)/\sqrt{1+n^3} & \sqrt{1+n^3}
\end{matrix}\right).
\end{equation}
Due to Eq. (\ref{DiagonalizedcolorDirectionMatrix}) one finds
\begin{equation}
g\,U E_n(t)\,U^\dagger = \frac{g}{2} \,\text{diag}\big(E(t),-E(t)\big)\,.
\end{equation}
This may be interpreted as effectively decomposing the $SU(2)$ gauge group into $U(1)\times U(1)$ with an Abelian coupling constant half as strong as the original non-Abelian one. By homogeneity and constancy of the transformation matrix, the given Abelianization procedure leaves equations of motion invariant.

\subsection{Correlation functions}
The time evolution of all fields involved allows the computation of a wide range of bosonic and fermionic observables at any simulation time. Inter alia, the color charge density $\rho_n^a(t)$ and the fermion color current $j_n^a(t)$ can be computed with the help of the statistical propagator (\ref{eq:statprop}) according to
\begin{widetext}
\begin{eqnarray}
\rho_n^a(t) & = & \frac{1}{2} \langle [\psi_n(t), \overline{\psi}_n(t) \gamma^0 T^a]\rangle = \text{Re}\,\text{Tr}\big[\Delta_{n,n} (t)\,\gamma^0 \,T^a\big],\\
j_n^a(t) &=&\frac{1}{2}\sum_{k=1}^{K}\sum_{m=1}^k C_k\langle  [\psi_{n+m}(t),\overline{\psi}_{n+m-k}(t)(\gamma^1-ikr)U_{n+m-k,k-m}T^a U_{n,m}]  \rangle\nonumber \\
&=&  \sum_{k=1}^{K} \sum_{m=1}^k C_k \,\text{Re}\, \text{Tr} \bigg[\Delta_{n+m,n+m-k}(t)\,\big(\gamma^1-ikr\big)\, U_{n+m-k,k-m}(t)\,T^a\, U_{n,m}(t)\bigg].\label{FermionicCurrentSpatial}
\end{eqnarray}
\end{widetext}
Here $j_n^a(t)$ naturally incorporates lattice improvements due to the spatial derivative in the respective continuum expression.

From $H^{(\text{g})}$ one infers the bosonic energy density at site $n\in\Lambda$,
\begin{equation}
\mathcal{E}_n^{(\text{g})}(t) = \frac{1}{2} \sum_{a=1}^3 E_n^a(t)\, E_n^a(t)\,.
\end{equation}
Though not indicated explicitly, correlation functions are computed as ensemble averages of the data calculated in individual runs.

\subsection{Abelianized fermion numbers}

For the interpretation of results, in particular in situations with homogeneous fields, it is often useful to define fermion pseudoparticle numbers from single-particle energy densities following the lines of Ref.~\cite{Kasper:2014uaa}. To this end, gauge degrees of freedom are considered to be spatially uniform at all times, i.e.~$A^a(t) \equiv A^a_n(t)$, $E^a(t)\equiv E^a_n(t)$. From the fermion Hamiltonian given in Eq.~(\ref{HamiltonianFermions}) we read off a Hamiltonian operator with spatial indices, $n,m\in\Lambda$:
\begin{align}
\mathcal{H}_{n,m}^{(\text{f})}(t) =\,\,&  \bigg(m+\frac{Kr}{a}\bigg)\,\delta_{n,m}\nonumber\\
& -\frac{1}{2a}\sum_{k=1}^{K}C_k\left(i\gamma^{1}+kr\right)U_{n,k}(t)\delta_{n+k,m} \nonumber\\
& + \frac{1}{2a}\sum_{k=1}^{K}C_k\left(i\gamma^{1}-kr\right)U_{n,-k}(t)\delta_{n-k,m}\,.
\end{align}
From this one can compute a fermion energy density as
\begin{equation}
\mathcal{E}^{(\text{f})}_n (t) = - \frac{1}{2} \sum_{m\in\Lambda}\, \text{Tr} \big(\mathcal{H}^{(\text{f})}_{n,m}(t)\, \Delta_{m,n}(t)\big)\,,
\end{equation}
with a trace over color and Dirac indices.
Using the diagonalization matrix $U$ as given in Eq.~(\ref{DiagonalizationMatrixU}) and
\begin{equation}
E^a(t) = n^a E(t),\,\,\,\,\,\,\,\,A_n^a(t) \equiv n^a A(t) = -n^a\int^t_0 dt' E(t'),
\end{equation}
one may diagonalize $\mathcal{H}_{m,n}^{(\text{f})}$,
\begin{eqnarray}
U\mathcal{H}_{n,m}^{(\text{f})}(t)\, U^\dagger &=& \bigg(m+\frac{Kr}{a}\bigg)\delta_{n,m}\nonumber\\
&&- \frac{1}{2a}\sum_{k=1}^K C_k (i \gamma^1+kr) T_k \delta_{n+k,m}\nonumber\\
&&+ \frac{1}{2a} \sum_{k=1}^K C_k (i \gamma^1-kr)T^\dagger_k\delta_{n-k,m}.
\end{eqnarray}
Here we employed
\begin{equation}
T_k\equiv \left(\begin{matrix}
e^{+ikga A(t)/2} & 0 \\
0 & e^{-ikga A(t)/2}
\end{matrix}\right).
\end{equation}
We define
\begin{subequations}
\begin{align}
\left(\begin{matrix}
\phi_{n,q}^{u/v,+}(t)\\
\phi_{n,q}^{u/v,-}(t)
\end{matrix}\right)
 &\equiv \, U
\left(\begin{matrix}
\phi_{n,q,1}^{u/v}(t)\\
\phi_{n,q,2}^{u/v}(t)
\end{matrix}\right)\,,\\
\mathcal{H}_{n,m}^{(\text{f}),+/-}(t)  \,&\equiv \, \big(U\mathcal{H}_{n,m}^{(\text{f})}(t)\, U^\dagger\big)_{11/22}\,.
\end{align}
\end{subequations}
From this we can compute a diagonalized phase-space energy density $\epsilon_{n,q}^\pm(t)$:
\begin{eqnarray}
\epsilon_{n,q}^{\pm}(t)&=&\frac{1}{2L^2} \sum_{m\in \Lambda}e^{2\pi i q_1 m/N_1} \sum_{q'\in \tilde{\Lambda}}\bigg[\overline{\phi}^{v,\pm}_{n,q'}\mathcal{H}_{n,m}^{(\text{f}),\pm}(t)\,\phi^{v,\pm}_{q,q'} \nonumber\\
&& - \overline{\phi}^{u,\pm}_{n,q'}\mathcal{H}_{n,m}^{(\text{f}),\pm}(t)\,\phi^{u,\pm}_{q,q'}\bigg]\, ,
\end{eqnarray}
where $\phi^{u/v,\pm}_{q,q'}$ denote the Fourier-transformed Abelianized mode functions,
\begin{equation}
\phi^{u/v,\pm}_{q,q'} \equiv a \sum_{n\in\Lambda} e^{-2\pi i n \tilde{q} / N_1} \phi^{u/v,\pm}_{n,q'}
\end{equation}
with $\tilde{q}\equiv q - N_1/2$.
We find an Abelianized single-particle energy density
\begin{equation}
\omega^\pm_{n,q}(t) \equiv \sqrt{\big(m^\pm_{n,q}(t)\big)^2 + \big(p^\pm_{n,q}(t)\big)^2},
\end{equation}
computed from physical masses $m^\pm_{n,q}(t)$ and physical momenta $p^\pm_{n,q}(t)$ given by
\begin{subequations}
\begin{align}
m^\pm_{n,q}(t) \equiv \,\,&  m + \frac{r}{2a}\sum_{k=1}^K k \, C_k \big[2 - V_{q,k}^\pm(t) - \overline{V}_{q,k}^\pm(t) \big]\,,\label{PhysicalMass}\\
p^\pm_{n,q}(t) \equiv \,\,& \frac{i}{2a} \sum_{k=1}^K C_k \big[\overline{V}_{q,k}^\pm (t) - V_{q,k}^\pm(t)\big]\label{PhysicalMomentum}\,.
\end{align}
\end{subequations}
For notational simplicity we introduced 
\begin{equation}
V_{q,k}^\pm(t) \equiv \exp\bigg[ik\bigg(\frac{2\pi \tilde{q}}{N_1} \pm \frac{gaA(t)}{2} \bigg)\bigg].
\end{equation}
Using $\omega^\pm_{n,q}(t)$ we may define an Abelianized fermion pseudoparticle number as follows:
\begin{equation}
n^\pm_{n,q}(t) \equiv\frac{\epsilon^\pm_{n,q}(t)}{2\omega^\pm_{n,q}(t)} + \frac{1}{2L}.
\end{equation}
From this we can compute marginal distributions via~\cite{Kasper:2014uaa}
\begin{subequations}
\begin{align}
n_{n}^\pm(t) &\equiv  \frac{1}{L} \sum_{q} n_{n,q}^\pm (t),\\
n_{q}^\pm (t) &\equiv a \sum_{n} n_{n,q}^\pm(t),
\end{align}
\end{subequations}
and a total pseudoparticle number as~\cite{Kasper:2014uaa}
\begin{equation}
n^\pm (t) = \frac{1}{N_1} \sum_{n,q} n_{n,q}^\pm (t).
\end{equation}
Since $n^\pm(t)$ represents an expectation value obtained from an ensemble average, it gives in general rise to noninteger particle numbers. In what follows we refer to this method of computing fermion numbers as Abelianized fermion numbers, indicating that the single-particle energy $\omega^\pm_{n,p}(t)$ has a well-motivated interpretation if one can diagonalize color degrees of freedom.

\subsection{Gauge-invariant fermion numbers}
For comparison, and for the interpretation of the results from inhomogeneous initial conditions, we consider a second definition of fermion number using a Wigner function approach \cite{Hebenstreit:2013qxa}. To this end, first a gauge-invariant statistical propagator is constructed,
\begin{equation}
\tilde{\Delta}_{m,n}(t) \equiv \text{tr}\big(W_{n,m}(t)\Delta_{m,n}(t)\big),
\end{equation}
with a trace over color indices only. $W_{n,m}(t)$ is constructed as the lattice Wilson line along the shortest spatial path that connects the points $n,m$. It transforms under a gauge transformation $V_n(t)\in SU(2)$ as $W_{n,m}(t)\mapsto V_n (t)W_{n,m}(t)V_m^\dagger(t)$, while $\Delta_{m,n}(t)\mapsto V_m (t)\Delta_{m,n}(t) V_n^\dagger(t)$. Setting $\delta \equiv n-m$ and dropping the time argument in the notation, if $\delta \geq 0$ we find
\begin{subequations}\label{eq:Wdef}
\begin{align}
\delta \leq \frac{N_1}{2}: &\,\,\,\,\,\, W_{n,m} \equiv U_{n-1}^\dagger U_{n-2}^\dagger\cdots U_{m+1}^\dagger U_m^\dagger\label{eq:Wdef:a}\\
\delta > \frac{N_1}{2}: &\,\,\,\,\,\, W_{n,m} \equiv U_n U_{n+1}\cdots U_{N_1-1} U_0 \cdots U_{m-2}U_{m-1}\,.\label{eq:Wdef:b}
\end{align}
If $\delta <0$:
\begin{align}
\delta > -\frac{N_1}{2}: &\,\,\,\,\,\, W_{n,m}\equiv U_n U_{n+1} \cdots U_{m-2}U_{m-1}\label{eq:Wdef:c}\\
\delta \leq -\frac{N_1}{2}: &\,\,\,\,\,\, W_{n,m}\equiv U_{n-1}^\dagger U_{n-2}^\dagger \cdots U_0^\dagger U_{N_1-1}^\dagger \cdots U_{m+1}^\dagger U_m^\dagger\,.\label{eq:Wdef:d}
\end{align}
\end{subequations}
From $\tilde{\Delta}_{n,m}(t)$ we compute the lattice Wigner function as
\begin{equation}
\mathcal{W}_{l,q}(t)\equiv -\frac{a}{2} e^{\pi i l q/N_q} \sum_{k\in \Lambda} e^{-2\pi i k q /N_1} \tilde{\Delta}_{k,[l-k]_{N_1}}(t) + \gamma.c.\, ,
\end{equation}
where $\mathcal{O}+\gamma.c.$ denotes $\mathcal{O}+\gamma^0 \mathcal{O}^\dagger \gamma^0$. Here, $l\in \Lambda_{\mathcal{W}}$ and $q\in \tilde{\Lambda}_{\mathcal{W}}$ with Wigner lattices
\begin{subequations}
\begin{align}
\Lambda_{\mathcal{W}} &= \bigg\{l=\frac{2x}{a}\in \{0,\dots, 2N_1-1\}\bigg\}\,,\\
\tilde{\Lambda}_{\mathcal{W}}  &= \bigg\{q = \frac{Lp}{\pi}\in \{-N_1,\dots, N_1-1\}\bigg\}\,.
\end{align}
\end{subequations}
We account for the periodicity of the lattice by taking the module operation in the second argument of $\tilde{\Delta}_{k,[l-k]_{N_1}}(t)$, with $[l-k]_{N_1}\equiv (l-k)\mod N_1$. Since the Wigner function $\mathcal{W}$ fulfils $\mathcal{W}^\dagger = \gamma^0 \mathcal{W} \gamma^0$, we can employ the decomposition \cite{Hebenstreit:2013qxa}
\begin{equation}
\mathcal{W} = \frac{1}{2} \left(\mathbbm{s} + i\gamma^5 \mathbbm{p} + \gamma^0 \mathbbm{v}_0 - \gamma^1 \mathbbm{v}\right),
\end{equation}
where $\mathbbm{s},\mathbbm{p},\mathbbm{v}_0,\mathbbm{v}$ are all real. In the free case, (\ref{PhysicalMass}) and (\ref{PhysicalMomentum}) reduce to
\begin{eqnarray}
m_q &\equiv &m + \frac{2r}{a}\sum_{k=1}^K k\, C_k \sin^2\bigg(\frac{k\pi \tilde{q}}{N_1}\bigg)\,,\label{FreePhysicalMass}\\
p_q &\equiv & \frac{1}{a}\sum_{k=1}^K C_k \sin \bigg(\frac{2k\pi \tilde{q}}{N_1}\bigg)\,.\label{FreePhysicalMomentum}
\end{eqnarray}
From the decomposition components we calculate various pseudodistributions:
\begin{eqnarray}
\varrho_{l,q}(t) &=  & g\mathbbm{v}_{0;l,q}(t)\,,\\
\epsilon_{l,q}(t) &=& m_q\, \mathbbm{s}_{l,q}(t) + p_q \,\mathbbm{v}_{l,q}(t)\,,
\end{eqnarray}
corresponding to charge and energy density, respectively. We define a phase-space resolved gauge-invariant fermion quasiparticle number:
\begin{subequations}
\begin{eqnarray}
n_{l,q} (t) \equiv \frac{\epsilon_{l,q}(t) - \epsilon_{\text{vac},l,q}(t) + \omega_{q} \mathbbm{v}_{0;l,q}(t)}{2\omega_q}\,,\\
\overline{n}_{l,q} (t) \equiv \frac{\epsilon_{l,q}(t) - \epsilon_{\text{vac},l,q}(t) - \omega_{q} \mathbbm{v}_{0;l,q}(t)}{2\omega_q}\,,
\end{eqnarray}
\end{subequations}
where $n_{l,q}(t), \overline{n}_{l,q}(t)$ label particle and antiparticle contributions, respectively. We will refer to this method of computing fermion numbers as gauge-invariant fermion numbers, justified by the included Wilson-line $W_{n,m}(t)$ to construct fermion numbers in a gauge-invariant fashion.

\subsection{Connected charge-charge correlation function}
In the study of color string dynamics a measure of connected color charge-color charge correlations is considered. For this purpose we introduce here the connected equal-time charge-charge correlation function
\begin{eqnarray}
C_{mn}^{ab}(t) &\equiv & \langle T \rho_m^a (t) \rho_n^b(t) \rangle - \langle \rho_m^a(t)\rangle \langle \rho_n^b(t)\rangle\nonumber\\
&=& \frac{1}{2}\langle \{ \rho_m^a (t), \rho_n^b(t)\} \rangle - \langle \rho_m^a(t)\rangle \langle \rho_n^b(t)\rangle.
\end{eqnarray}
In terms of mode functions one finds
\begin{eqnarray}\label{ConnectedChargeChargeCorrelFunctionModes}
C_{mn}^{ab}(t) &=& \frac{1}{2L^2} \sum_{q,p\in \tilde{\Lambda}}\sum_{c,d^{\mathstrut}=1}^2 \bigg[ \phi^{v,\dagger}_{m,q,c} T^a \phi^{u\mathstrut}_{m,p,d}\phi^{u,\dagger}_{n,p,d} T^b \phi^{v\mathstrut}_{n,q,c} \nonumber\\
&& +  \phi^{u,\dagger}_{m,q,c} T^a \phi^{v\mathstrut}_{m,p,d}\phi^{v,\dagger}_{n,p,d} T^b \phi^{u\mathstrut}_{n,q,c}\bigg].
\end{eqnarray}

\subsection{Gauss's law}
To simulate states in the physical Hilbert space, Gauss's law needs to be met. In our model it reads
\begin{equation}\label{GaussLaw}
G_n^a \equiv  \frac{1}{a} \big(E_n^a - (U^\dagger_{n-1}E^{\mathstrut}_{n-1}U^{\mathstrut}_{n-1})^a\big) = -\rho_n^a.
\end{equation}
If the color charge $\rho_n^a$ vanishes, we can iteratively solve this equation for arbitrary $n>n_{\text{ref}}\in\Lambda$:
\begin{eqnarray}
E_n^a(t) &=& \big(U_{n-1}^\dagger(t)\,U_{n-2}^\dagger(t)\cdots U_{n_{\text{ref}}}^\dagger(t)\, E_{n_{\text{ref}}}^{\mathstrut} (t)\,U_{n_{\text{ref}}}^{\mathstrut}(t)\nonumber\\
&& \cdots U_{n-2}^{\mathstrut}(t)\,U_{n-1}^{\mathstrut}(t)\big)^a.
\end{eqnarray}
For the $G_n^a$ commuting with the model's Hamiltonian $H$, a state that initially fulfils Gauss's law does so for an arbitrary later point in time as well.

\section{Fermion production and improved Hamiltonian benchmarks}\label{SectionHomogeneous}
In this section we investigate fermion production from homogeneous chromoelectric fields exceeding the critical field strength for Schwinger pair creation. In particular, we establish how improved Hamiltonians can be used to simulate this process and emergent phenomena such as plasma oscillations using significantly smaller lattices as compared to unimproved lattice implementations.

\subsection{Fermion production: Benchmark at early times}

To benchmark the lattice setup, we first compare our lattice simulation results with established analytic formulas for the non-Abelian Schwinger mechanism in a static coherent chromoelectric background field~\cite{Gelis:2015kya, Nayak:2005pf, Hebenstreit:2010vz}. With this in mind we disregard the backaction of the fermion sector onto the gauge sector and exclude sampling. This approximate description is expected to be accurate at sufficiently early times, for which the analytic results are available. To be able to go to later times, we consider the fully nonlinear dynamics including backaction below. 

We introduce the dimensionless field-strength parameter
\begin{equation}
\epsilon^a = \frac{E_0^a}{E_c}\,,
\end{equation}
with $E_c = m^2/g$ being the critical field strength. We fix $g/m=0.1$ throughout this subsection and apply vacuum initial conditions for fermions, while the homogeneous chromoelectric field is set to the value stated. The simulations correspond to solving the fermion equation of motion, Eq. (\ref{EOMFermionModes}), with a sudden switch-on of the chromoelectric field at initial time $t_0 =0/m$.

\begin{figure}
    \centering
	\includegraphics[scale = 0.87]{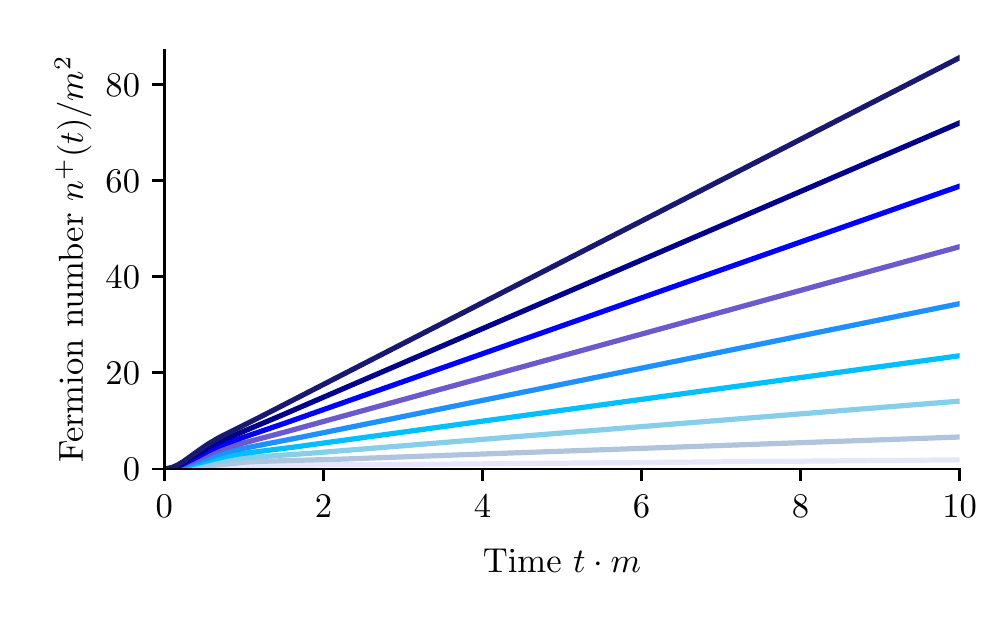}
	\caption{Abelianized fermion numbers from simulations for various background field strength $\epsilon=(\epsilon^1,0,0)$, ranging from $\epsilon^1=10$ (dark blue line) to $\epsilon^1=2$ (light blue line) in integer steps. The lattice parameters are $N_1=512$, $L=25.6/m$, $a=0.05/m$, $a_t=0.05 \,a$, computed at NLO.}\label{FigFermionNumbersVsTime}
\end{figure}

In Fig.~\ref{FigFermionNumbersVsTime} the total number of fermions as a function of time for diagonalized color direction + and various background fields $\epsilon = (\epsilon^1,0,0)$ is displayed. The calculations are based on NLO lattice improvements with lattice parameters as given in the figure caption. After a transient regime of enhanced fermion production at small times $t_{\textup{tr}}\lesssim 1/m$, the curves show a linear regime in which fermion-antifermion pairs are produced at a constant rate.

Linearly fitting the data curves for $n^{+}(t)/m^2$ at times $t\geq t_{\text{tr}}$, the rates together with results from respective analytical calculations are displayed in Fig.~\ref{FigFermionNumbersVsField}. As detailed in Appendix \ref{AppendixAnalytics}, one expects a fermion production rate per unit length and per diagonalized color direction of
\begin{equation}
\frac{\dot{n}^\pm}{Lm^2} = \frac{\epsilon}{4\pi} \exp\bigg(-\frac{2\pi}{\epsilon}\bigg)\,.
\end{equation}
The results from simulations are seen to agree with the analytics to very good accuracy. An exponential suppression of pair production below $E_c$ takes place both analytically and numerically, in accordance with prior studies in the framework of the Schwinger model and QED~\cite{Tanji:2010eu, Hebenstreit:2013qxa, Kasper:2014uaa}.

\begin{figure}
    \centering
	\includegraphics[scale = 0.87]{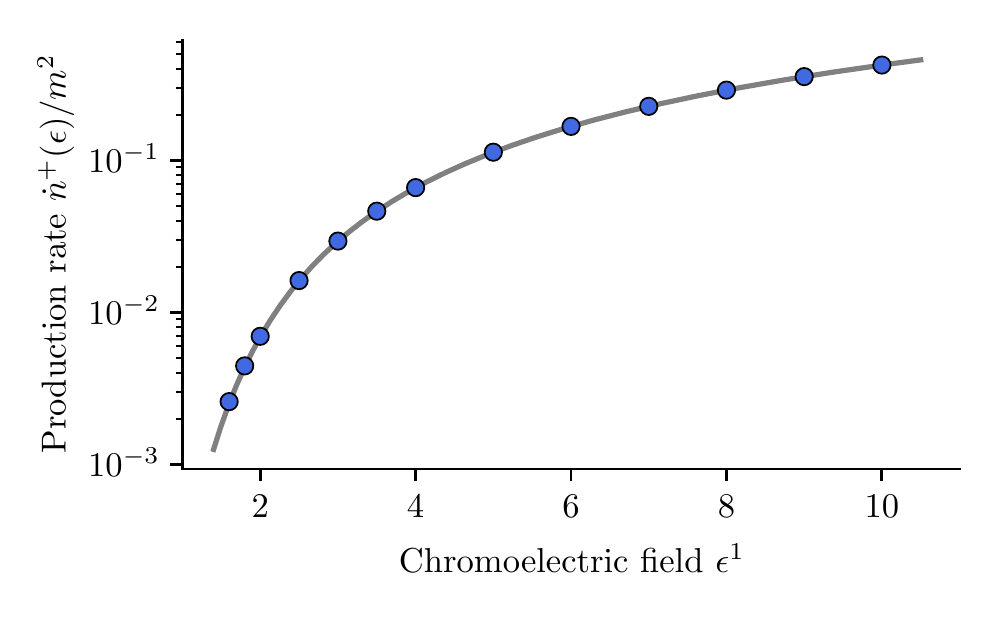}
	\caption{Abelianized fermion production rates from both simulations (blue dots) and analytics (solid, gray line). The employed lattice parameters are the same as for Fig.~\ref{FigFermionNumbersVsTime}. The background chromoelectric field is $\epsilon=(\epsilon^1,0,0)$, with $\epsilon^1$ being varied here.}\label{FigFermionNumbersVsField}
\end{figure}

\begin{figure}
    \centering
	\includegraphics[scale = 0.87]{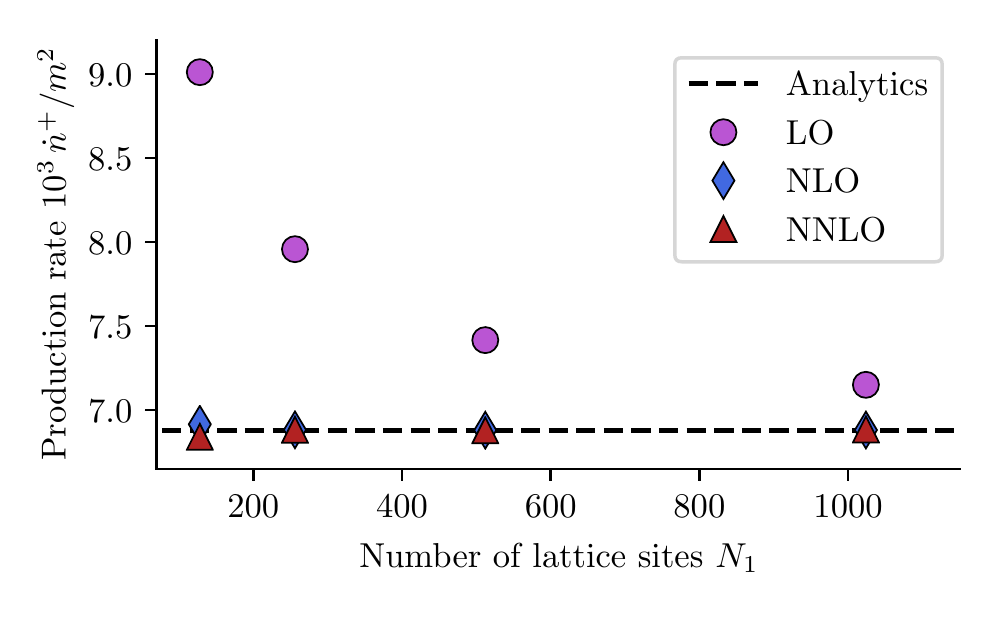}
	\caption{Comparison of the Abelianized fermion production rate from simulations with the analytic one-loop result. Shown are results without lattice improvements (violet dots),  NLO improvements (blue diamonds),  NNLO improvements (red triangles) and the corresponding analytical result (black dashed line). The lattice parameters are $L=25.6/m$, $a=L/N_1$, $a_t=0.02 \,a$ for a background chromoelectric field $\epsilon=(2,0,0)$.}\label{FigProductionRatesConvergence}
\end{figure}

We now study the role of lattice improvements for the approach of the simulation results of the discretized theory in a finite volume to the analytic prediction for the continuum theory in an infinite volume. Figure~\ref{FigProductionRatesConvergence} shows results for fermion production rates for various numbers of lattice sites $N_1$ in the unimproved LO formulation, which is then compared to NLO and NNLO improvements. Here, the total lattice size $L=N_1a$ is kept constant such that $a$ is decreasing with increased $N_1$ to approach the continuum.

For the unimproved lattice theory we find that getting close to continuum results for the production rate requires extremely small lattice spacings, in accordance with previous studies in QED$_{1+1}$~\cite{Hebenstreit:2013qxa}. We observe that this situation changes dramatically, once improved Hamiltonians are employed. Figure~\ref{FigProductionRatesConvergence} indicates that significantly smaller numbers of lattice sites $N_1$ lead already to results in the vicinity of the continuum theory. The NLO improved theory is seen to be extremely efficient with only minor differences to NNLO results, both converging very well at least for $N_1 \gtrsim 250$ for the employed parameters. Already around $N_1 \simeq 100$ the NLO (NNLO) improvement only slightly overestimates (underestimates) the continuum result, and smaller lattices may be employed if one is willing to accept errors exceeding the few-percent level. Relatively small lattices are crucial, e.g., for any realistic chance to implement this physics in quantum simulators in the not too distant future.  

\begin{figure}
    \centering
	\includegraphics[scale = 0.87]{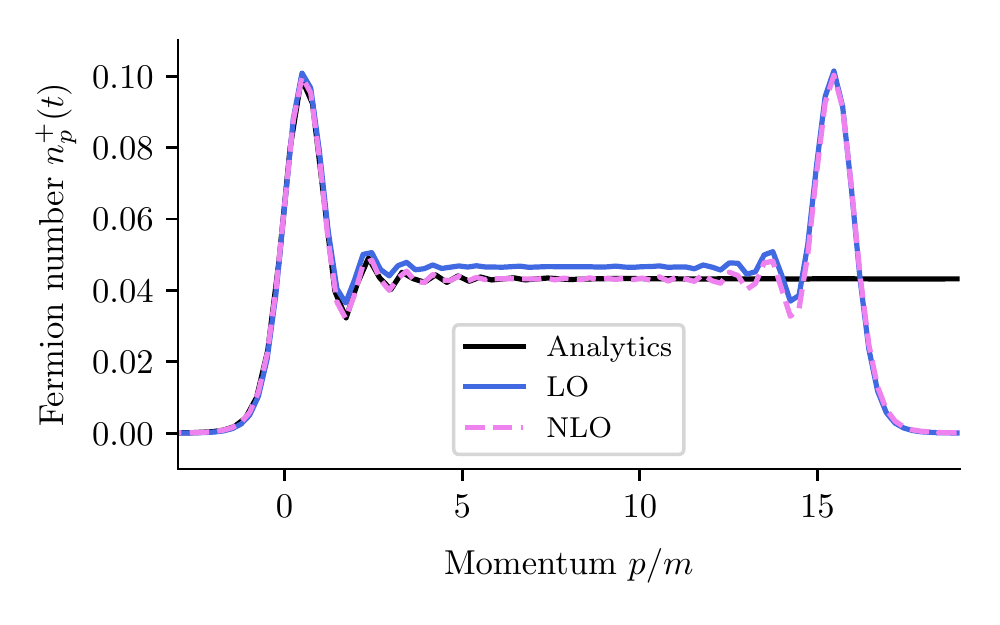}
	\caption{The Abelianized fermion number momentum spectrum without (LO, solid, blue line) and with first-order (NLO, dashed, pink line) lattice improvements, as well as analytics (solid, black line). All curves show the diagonalized '$+$' color at simulation time $t=16/m$. The employed lattice parameters read $N_1=512$, $L=25.6/m$, $a=0.05/m$, $\epsilon=(2,0,0)$, $a_t=0.05 \,a$.}\label{FigFermionNumbersVsMomentum}
\end{figure}

Apart from integrated quantities such as the total particle number, it is instructive to analyze also momentum-resolved fermion numbers. In Fig.~\ref{FigFermionNumbersVsMomentum} the momentum spectrum of created fermions, $n_p^+(t)$, is displayed at time $t = 16/m$ for diagonalized color direction +. Switching to diagonalized color direction - is equivalent to reflecting the graph at $p/m=0$. Shown are results without (LO) and with first-order (NLO) lattice improvements for a fixed number of lattice sites as given in the figure caption, together with the analytic result. 

The fermion number distribution shows a clear peak around $p/m=0$, reflecting the fact that most of the fermion-antifermion pairs are created at rest. Subsequently, they are accelerated in the applied chromoelectric field towards higher momenta. The low-momentum fermions are seen to be rather well described both at LO and NLO. At higher momenta, however, the accelerated fermions are significantly better described using the lattice improved Hamiltonian. These improvements become particularly visible for integrated quantities such as the total particle number, which sum over all momentum modes.

We note that in analytical computations the initial time is sent to the remote past, $t_0\to -\infty$, such that produced particles occupy arbitrarily high momenta. In contrast, the actual simulations start at $t_0=0/m$ and produced particles only occupy finite momenta at finite times in the presence of a constant background field. This initial-time difference is also the reason for the transient regime of enhanced fermion production at small times $t_{\textup{tr}}\lesssim 1/m$ visible in the simulation data of Fig.~\ref{FigFermionNumbersVsTime}, which is not present in the analytic estimates. Of course, only at sufficiently early times restricting to a constant background field is a valid approximation. Since total energy must be conserved, at later times the backaction of the produced fermion pairs on the applied chromoelectric field becomes relevant, which we address in the following. 

\subsection{Plasma oscillations: reaching longer times using lattice improved Hamiltonians}\label{SecPlasmaOscillations}
We now include the back-coupling of fermion currents onto the gauge sector, correspondingly taking the chromoelectric field equation of motion (\ref{EOMChromoelectricField}) into account. We keep $g/m=0.3$ fixed throughout this subsection.

\begin{figure}
    \centering
	\includegraphics[scale = 0.87]{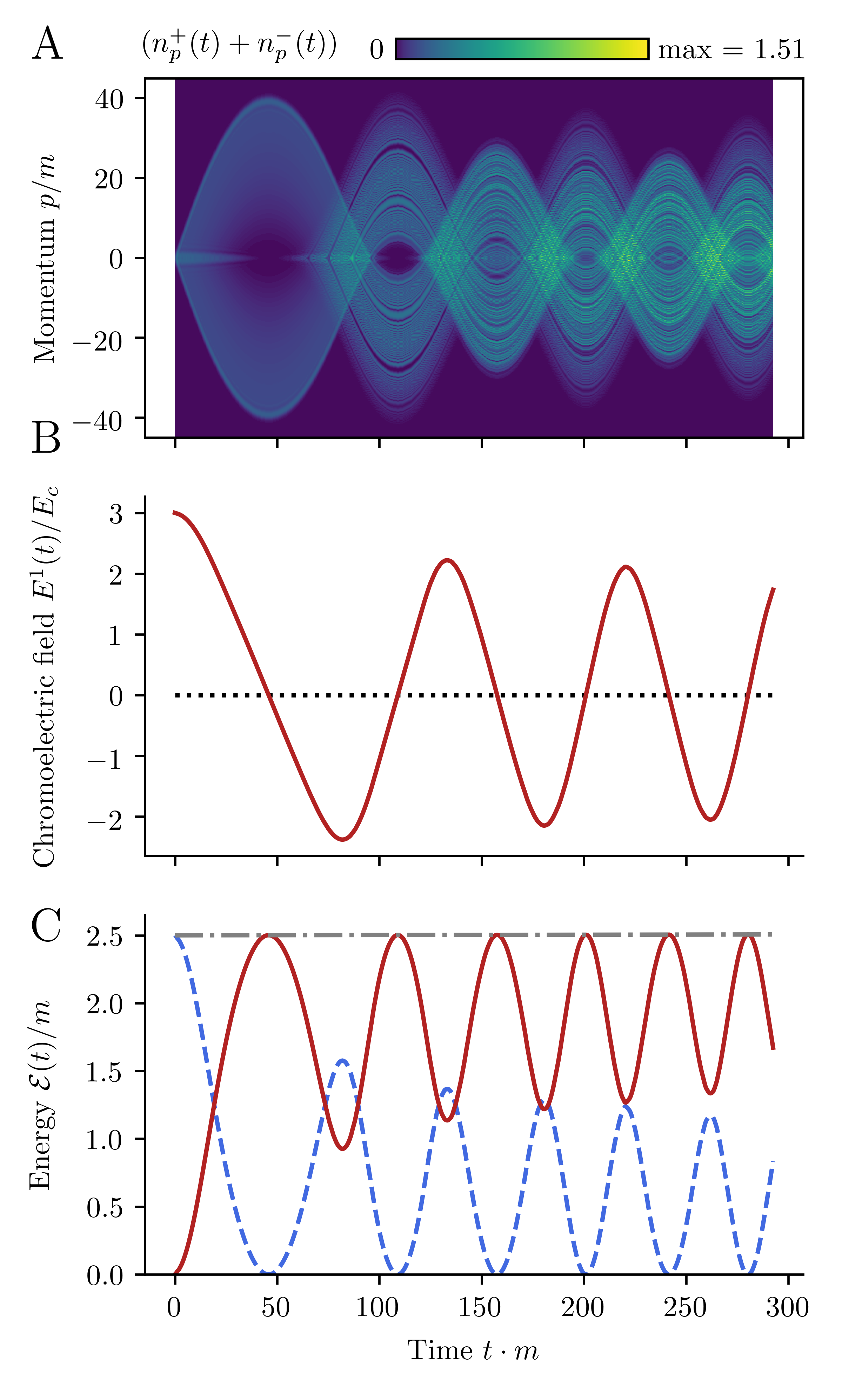}
	\caption{Plasma oscillations manifest in multiple observables. \textbf{Panel A}: Fermion number momentum spectrum $n_p^{+}(t)+n_p^{-}(t)$. \textbf{Panel B}: Volume-averaged chromoelectric field $E^1(t)$, $E^2(t)=E^3(t)=0$. \textbf{Panel C}: Gauge field energy (blue, dashed line), fermion energy (red, solid line) and total energy (grey, dotted line) with the fermion vacuum energy subtracted from fermion and total energy. Lattice parameters of all panels are $N_1=768$, $L=38.4/m$, $a=0.05/m$, $a_t=0.02 \,a$, $g/m=0.3$, computed at NNLO, with initial chromoelectric field $\epsilon = (3,0,0)$.}\label{FigPlasmaOscillations}
\end{figure}

Fig.~\ref{FigPlasmaOscillations}A displays the Abelianized fermion number momentum spectrum $n_p^{+}(t)+n_p^{-}(t)$ as a function of time. The initial acceleration of the produced fermions is visible along with a subsequent deceleration process, then an acceleration to lower maximum momenta than before and so on. To understand this oscillating behavior, it is helpful to consider also the time evolution of the chromoelectric field displayed in Fig.~\ref{FigPlasmaOscillations}B. A fermion current is induced at times when the gauge field is strong, accompanied by a respective gauge field decay. Once the gauge field decayed fully, via the produced fermions' backaction onto the gauge sector a gauge field builds up again but pointing in the opposite direction. This process occurs again and again, resulting in non-Abelian plasma oscillations. 

In Fig.~\ref{FigPlasmaOscillations}C different contributions to the total energy density are shown as a function of time. The gauge part reflects well the oscillating behavior of the chromoelectric field. The fermion energy part is seen to be large whenever the chromoelectric field part is small. Their sum stays constant, as required by energy conservation. 

All results shown in Fig.~\ref{FigPlasmaOscillations} have been obtained at NNLO. We now address the role of lattice improvements for the nonlinear dynamics. We have seen that the back-coupling of fermions and gauge fields leads to plasma oscillations with decreasing maximum fermion momenta and multiple zero crossings at later times. In general, finite-volume effects and the associated limits on resolving low-momentum scales are expected to play an enhanced role at late times. 

\begin{figure}
    \centering
	\includegraphics[scale = 0.87]{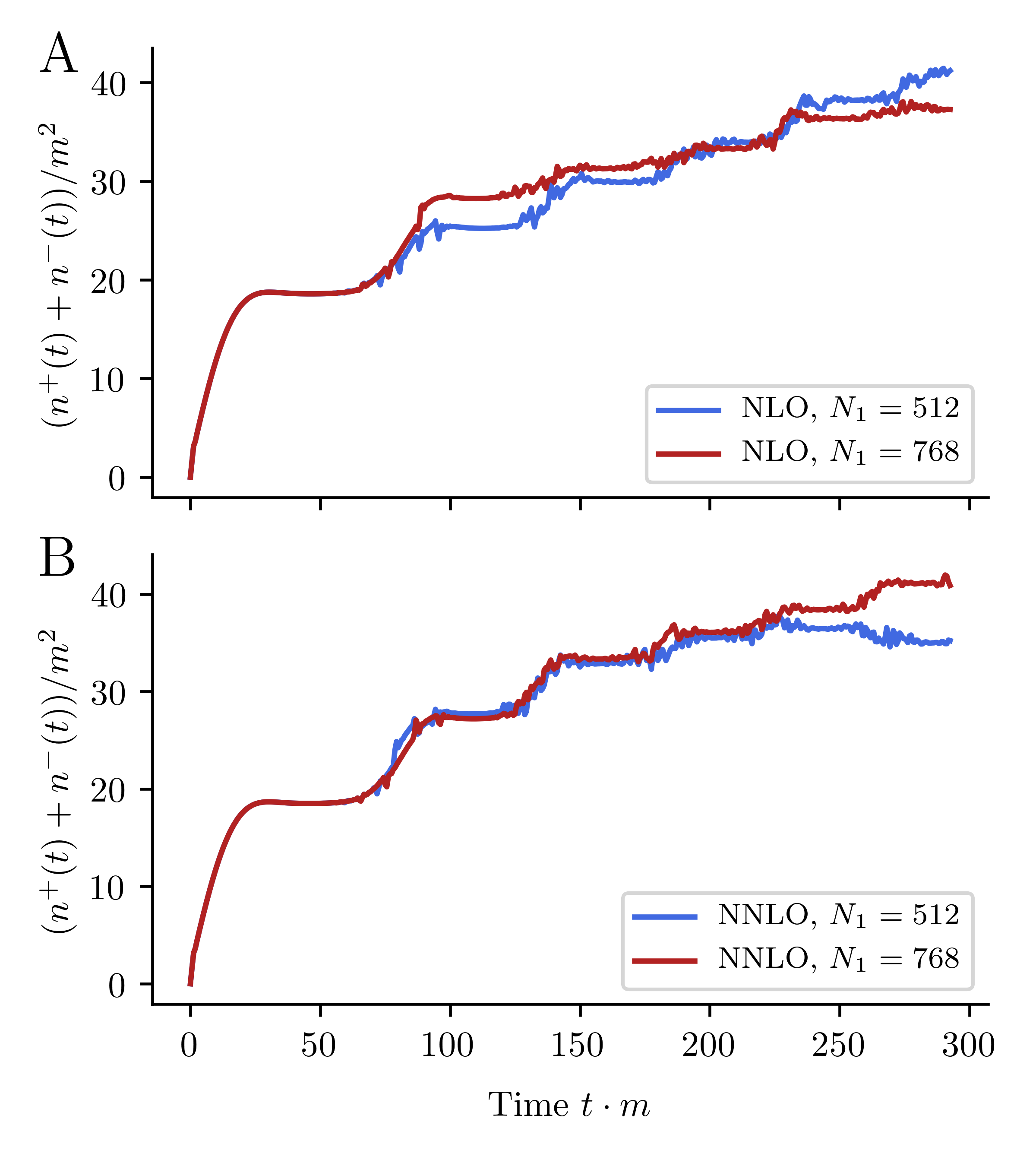}
	\caption{Total fermion numbers $n^{+}(t)+n^{-}(t)$ versus time, with $N_1=512$, $L=25.6/m$ (blue line) and $N_1 = 768$, $L=38.4/m$ (red line). \textbf{Panel A}: NLO improvements. \textbf{Panel B}: NNLO improvements. The remaining parameters of both panels are as in Fig.~\ref{FigPlasmaOscillations}.}\label{FigFermionNumberContinuumLimitImprovements}
\end{figure}

To illustrate the dependence of the results on the volume, we display in Fig.~\ref{FigFermionNumberContinuumLimitImprovements} the total fermion number $n^{+}(t)+n^{-}(t)$ as a function of time. The computations are done for two different lattice sizes with $N_1 = 512$ and $N_2 = 768$ for fixed lattice spacing such that the volumes differ accordingly. Panel A displays respective results at NLO, for which one observes a practically volume-independent early-time behavior. But already after about the first plasma oscillation, indicated by a plateau in the function $n^{+}(t)+n^{-}(t)$ which is due to a zero-crossing of the chromoelectric field around that time, deviations between the $N_1 = 512$ and $N_2 = 768$ setups become visible. This has to be confronted with the corresponding simulations at NNLO given in panel B. In this case, both lattice sizes give very similar results up to about the time of the fourth plateau, and even afterwards the larger lattice seems to give reasonable predictions. The ability to quantitatively describe longer time scales for a given lattice size is a very powerful property of the lattice improvements.

We end this section by noting that the qualitative behavior of the plasma oscillations in QCD$_{1+1}$, including the growth in oscillation frequency with time, coincides well with previous classical-statistical studies in the framework of $U(1)$ gauge theory \cite{Hebenstreit:2013qxa, Kasper:2014uaa}. However, from Fig.~\ref{FigPlasmaOscillations}A a characteristic property of $SU(2)$ theory plasma oscillations can be observed: Against the gauge field background half of the produced fermions are accelerated into positive direction, while simultaneously the other half is accelerated into negative spatial direction. The gauge field changing sign around its first 0, the produced fermions start being accelerated into the respectively opposite direction. The pattern of fermion acceleration closely mimics the oscillating gauge field dynamics at later times, too.

Resulting from temporary fermion production dropouts in the zero-momentum mode, substructures emerge in the time-evolving momentum spectrum of fermions from times $t\simeq 80/m$ onwards. In fact, the pattern is a finite-volume lattice artefact, continuously vanishing with an increasing number of lattice sites. For corresponding details we refer to Appendix~\ref{AppendixInfraredArtefacts}.

\begin{figure}
    \centering
	\includegraphics[scale = 0.87]{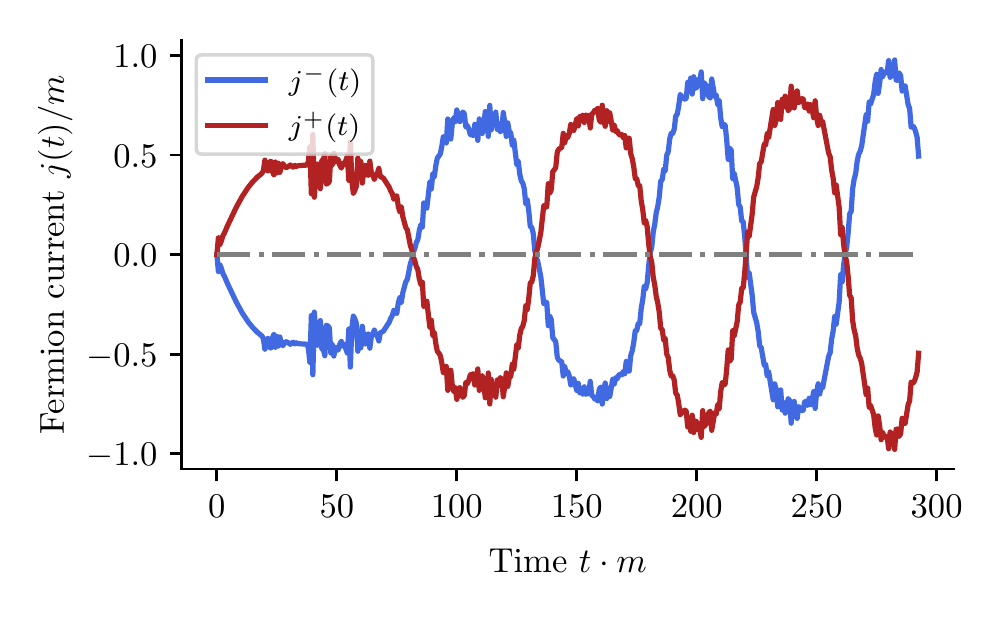}
	\caption{Volume-averaged Abelianized currents $j^{\pm}(t)$. The initial chromoelectric field is  $\epsilon = (3,0,0)$. Displayed are both the diagonalized + color direction (red line) and the -- direction (blue line), as well as their sum (grey, dotted line). Parameters are the same as in Fig. \ref{FigPlasmaOscillations}.}\label{FigOscillatingCurrentsVsTime}
\end{figure}
For homogeneous initial conditions we can diagonalize color degrees of freedom as pointed out in Sec.~\ref{SubsectionAbelianization}. We define the Abelianized fermion current using the already established fermion current (\ref{FermionicCurrentSpatial}) and the diagonalization matrix $U$, given by Eq.~(\ref{DiagonalizationMatrixU}),
\begin{equation}
\left(\begin{matrix}
j^+_n(t) & 0\\
0 & j^-_n(t)
\end{matrix}\right) \equiv \sum_{a=1}^3 U j^a_n(t) T^a U^\dagger.
\end{equation}

In Fig.~\ref{FigOscillatingCurrentsVsTime} the Abelianized fermion currents $j^{\pm}(t)$ are shown. One being the negative of the other, they indicate propagation of the produced fermions into opposite spatial directions. The oscillating behavior is a manifestation of the occurring plasma oscillations. The two Abelianized fermion currents sum up to 0, in accordance with both theory and previous studies~\cite{Tanji:2010eu, Tanji:2015ata}.

\section{String-breaking dynamics and higher correlation functions}\label{SectionInhomogeneous}

After the benchmark tests for homogeneous fields in the previous section, we now apply the lattice improved Hamiltonian approach to string-breaking dynamics in QCD$_{1+1}$. This involves computations of the nonequilibrium dynamics for inhomogeneous field configurations, constituting a particular strength of the approach. Moreover, we demonstrate that even higher-order correlation functions such as the charge-charge correlator (\ref{ConnectedChargeChargeCorrelFunctionModes}) involving four fermion fields are accessible with these techniques.

\subsection{Setup and initial conditions}

To motivate our setup and initial conditions, we start by considering 
a confining gauge string of length $d = la$ between two external color charges, specified classically by the color charge distribution
\begin{equation}\label{StringInitialColorCharge}
\rho_{0,n}^3 = gN_0^3\big(\delta_{n,(N_1+l)/2} - \delta_{n,(N_1-l)/2}\big),
\end{equation}
whereas $\rho_{0,n}^1 = \rho_{0,n}^2 = 0$ due to $N_0^1=N_0^2 = 0$. By Gauss's law (\ref{GaussLaw}) this results, classically, in a homogeneous chromoelectric field 3-component along the string,
\begin{equation}
E_n^3 = \left\{
\begin{array}{ll}
gN_0^3 & \textup{if }(N_1-l)/2 \leq n < (N_1+l)/2,\\
0 & \textup{else}.
\end{array}
\right.
\end{equation}
The string constructed in such a way from external color charges $\pm gN_0^3$ has an energy content of
\begin{equation}
V_{\textup{str}} = \frac{g^2 (N_0^3)^2 d}{2}.
\end{equation}
If we demand that the initial classical chromoelectric field in the string's interior reads by value $\epsilon_{\textup{init}}=(0,0,c)$ for a real constant $c$, the external color charges need to provide the energy
\begin{equation}
V_{\textup{str}} = a\sum_{n\in\Lambda} \mathcal{E}^{(\textup{g})}_n = \frac{c^2 E_c^2 d}{2}.
\end{equation}
This results in the requirement
\begin{equation}
\frac{g}{m} = \sqrt{\frac{c}{N_0^3}}.
\end{equation}
We apply this argumentation to a string with initial interior chromoelectric field $\epsilon_{\textup{ini}}=(0,0,2.0)$. Thus obtaining $g=m$ for $N_0^3 = 2.0$, we keep $g=m$ fixed throughout this section. Simulations to be described in the following demonstrate that to show critical string-breaking behavior the string length needs to be around $d_{\textup{crit}}^{SU(2)} \simeq 40.5/m$. Comparing to the critical string in $U(1)$-theory with $d_{\textup{crit}}^{U(1)} \simeq 28/m$ \cite{Hebenstreit:2013baa}, from 
\begin{equation}
40.5/m\simeq \sqrt{2} d_{\textup{crit}}^{U(1)} = \sqrt{N_0^{U(1)}} d_{\textup{crit}}^{U(1)}
\end{equation}
we deduce that the critical $SU(2)$-string  in QCD$_{1+1}$ corresponds to an Abelian string in $1+1$ dimensions with effective external $U(1)$-charges~\cite{Hebenstreit:2013baa, Kasper:2016mzj} of $N_0^{U(1)} = 2$.

These initial conditions are supplemented by vacuum fluctuations in the gauge sector. As in previous sections, for fermions we employ vacuum initial conditions. We discuss the impact of initial vacuum fluctuations on the gauge fields below and for details on their implementation we refer to Appendix \ref{AppendixBosonicQuantumFluctuations}, again. The number of averaged runs is set to 3 throughout this section, whenever classical-statistical sampling for the generation of the initial vacuum ``quantum-half'' for gauge fields is applied.

\subsection{String breaking and supercritical color strings}\label{SectionStringBreaking}

A string between two external charges that breaks completely, in particular with the center chromoelectric field asymptotically approaching 0, is called a critical string. A ``supercritical'' string is a string that shows multiple string breaking~\cite{Hebenstreit:2013baa}, such that the center chromoelectric field oscillates around 0 as it does in plasma oscillations.

\begin{figure*}
    \centering
	\includegraphics[scale = 0.87]{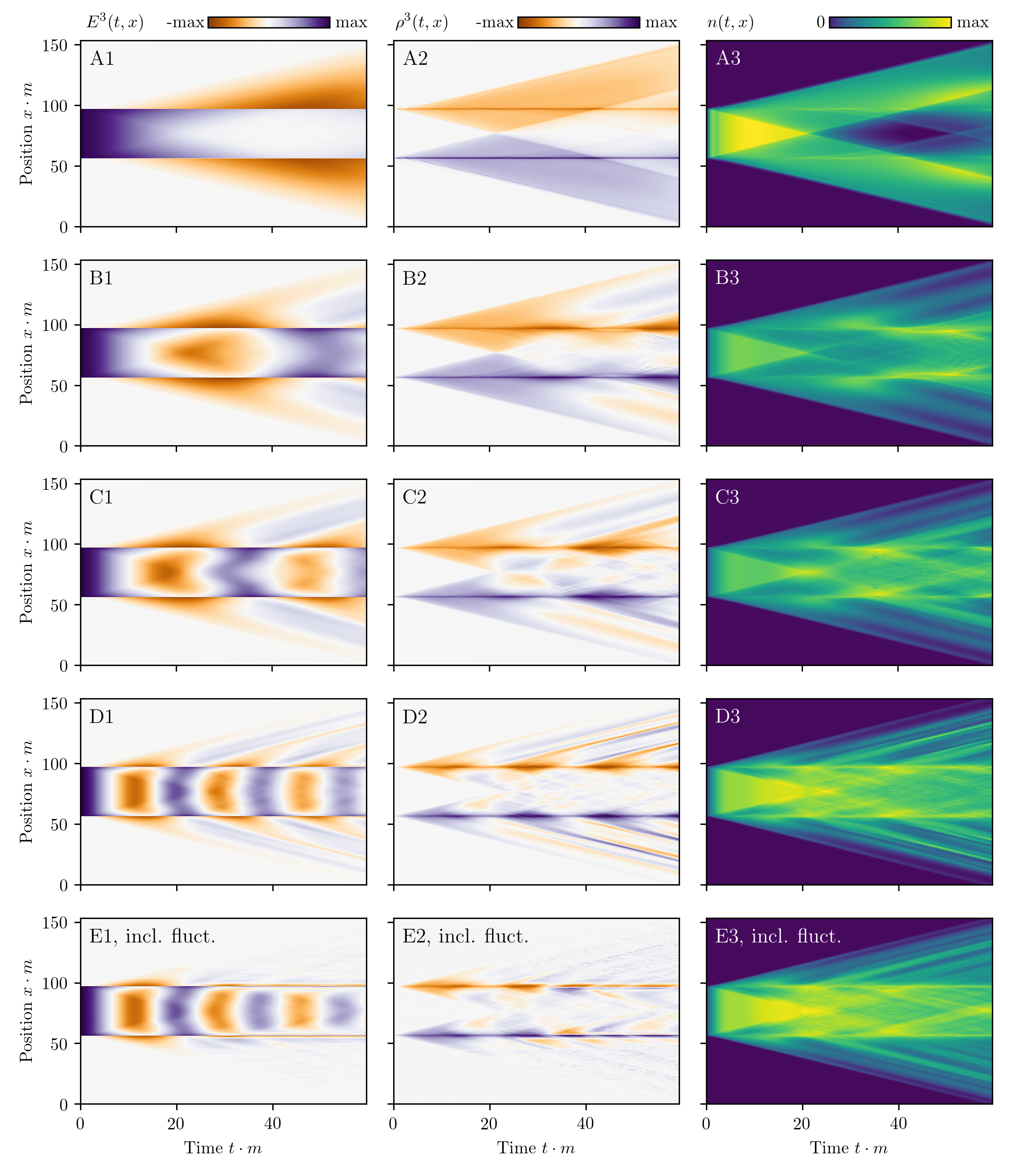}
	\caption{Color string characteristics for various external color charges. The color map of each graph is adjusted to its individual maximum. \textbf{Column 1}: Chromoelectric field 3-component $E^3(t,x)$. \textbf{Column 2}: Color charge distribution 3-component $\rho^3(t,x)$. \textbf{Column 3}: Gauge-invariant fermion numbers. \textbf{Row A}: External charges $N_0^3=2.0$. \textbf{Row B}: External charges $N_0^3=3.0$. \textbf{Row C}: External charges $N_0^3=4.0$. \textbf{Row D}: External charges $N_0^3=8.0$. \textbf{Row E}: External charges $N_0^3=8.0$, including vacuum gauge field fluctuations, 3 sampling runs averaged. Parameters of all panels are $N_1=1536$, $L=153.6/m$, $a=0.1/m$, $a_t = 0.02\, a$, $g/m=1$, NLO improvements, string length $d=d_{\textup{crit}}^{SU(2)} = 40.5/m$.}\label{FigStrings}	
\end{figure*}

Illustrating string-breaking dynamics by means of the gauge field, the color charge distribution and fermion numbers, in Fig.~\ref{FigStrings} we display space- and time-resolved simulational outcomes for four values of external charges $N_0^3 = 2.0,3.0,4.0,8.0$. In the absence of initial quantum fluctuations in the gauge field sector, due to $N_0^1=N_0^2 = 0$ all dynamics take place in the 3-components; 1- and 2-components stay 0 at all times simulated. If vacuum gauge field fluctuations are taken into account for the present initial conditions, this remains approximately true. The gauge-invariant fermion number definition is employed. For a comparison with Abelianized fermion numbers we refer to Appendix \ref{AppendixCompareFermionNumbersDefs}, the essence being that the two agree well with each other.

In row A of Fig.~\ref{FigStrings} variables are displayed for external charges $\pm 2.0 g$. At early times, fermion-antifermion pairs are produced via the non-Abelian Schwinger mechanism, as can be seen in fermion numbers (panel A3). The dynamically created fermions and antifermions are initially produced on top of each other, resulting in the absence of color charge inside the string at early times (panel A2). They get separated with nearly the speed of light by the chromoelectric field, gradually screening the external color charges. First, the center chromoelectric field roughly decays linearly, later asymptotically approaching zero (panel A1). Indeed, we find that for a string of length $d=40.5/m$ complete string breaking occurs. Additionally, part of the produced particles gets accelerated towards the outside of the string and propagates freely with approximately the speed of light beyond the external charges at both string ends, continuously occupying the space that surrounds the string, as can be observed in all three variables (panels A1-A3). Once the string broke and the external charges got screened, the sole dynamical objects that remain --- apart from small effects close to and inside the string --- are these fermions and antifermions that at constant speed fly away from the string. This behavior matches closely the corresponding Schwinger model string-breaking behavior \cite{Hebenstreit:2013baa, Kasper:2016mzj}.

While in row A of Fig.~\ref{FigStrings} the color string breaks exactly once, in rows B to E the strings break multiple times. The external charges are set to $\pm 3.0 g$, $\pm 4.0 g$ and $\pm 8.0 g$, giving rise to supercritical color strings. Due to the stronger initial chromoelectric field the external charges get screened faster than in the $N_0^3=2.0$-case. The strings break faster. In all four cases, however, the center chromoelectric field does not asymptotically approach 0 within the simulated time interval, but instead shows an oscillating behavior, indicating the occurrence of plasma oscillations and multiple string breaking. In contrast to the homogeneous plasma oscillations described in Sec.~\ref{SecPlasmaOscillations}, the plasma oscillations inside color strings happen inhomogeneously (panels B1-E1). By Gauss's law (\ref{GaussLaw}) chromoelectric field inhomogeneities are generated by the presence of color charges. Considering color charge distributions, we find a clear pattern of peaks and falls that oscillates in time (panels B2 to E2), getting more fragmented with increasing time the higher the external charge values get. Within this process, color charge accumulations occur that by sign differ from their immediate neighborhood. As in the critical case, part of the produced fermions and antifermions is accelerated towards the outside of the supercritical string, propagating approximately freely beyond the external charges at both string ends. But now the color charge outside the string varies strongly both spatially and temporally. Moreover, the space- and time-resolved fermion numbers closely follow the chromoelectric field dynamics. A large amount of fermion-antifermion pairs is produced wherever the gauge field is large (panels B3-E3). This is in accordance with the non-Abelian Schwinger mechanism.

We now address the role of vacuum fluctuations in the initial conditions of the gauge field configurations considered. In fact, the data shown in rows A to D are obtained without initializing the vacuum quantum-half for gauge field configurations. The initial vacuum fluctuations translate for each sample run into additional small inhomogeneities, which are averaged in the end. We check in the following whether this could change some of the smaller substructures observed, in particular in the $N_0^3=8.0$-case. 
The simulations for rows D and E of Fig.~\ref{FigStrings} are both generated with external charge values of $\pm 8.0 g$, in row E including vacuum fluctuations in gauge field initial conditions for comparison. While qualitatively both D and E show very similar results, in the chromoelectric field and color charge distributions tiny spatiotemporal oscillations are present if fluctuations are taken into account (panels D1-E2). Fermion numbers are nearly not affected by these oscillations since fermion production by small chromoelectric fields is strongly suppressed in the non-Abelian Schwinger mechanism (panel E3). We expect the tiny oscillations to become smaller with an increasing number of samplings. Furthermore, the behavior of chromoelectric fields in the immediate vicinity of the external charges exhibits some differences: While without initial gauge field quantum fluctuations the chromoelectric field outside the string shows distinctive falls approximately at times $15/m$, $33/m$ and $50/m$ and peaks at times $22/m$, $39/m$ and $57/m$ inside the string (panel D1), the last two peaks and falls both inside and outside the string in the external charges' vicinity nearly disappear if fluctuations are taken into account (panel E1). Thus, taking into account the initial vacuum gauge field fluctuations by means of sampling can have an effect on the oscillating chromoelectric field ``afterglow'' around external charges sitting at color string ends. The color charge distribution resembles this effect (panels D2 and E2); so does the fermion number distribution, though smaller by value (panels D3 and E3). 

Fundamentally, the external charges at supercritical color string ends never get screened fully: Always a remnant chromoelectric field inside the string is present. Time-evolving the given initial conditions, supercritical color strings remain confining in our model at all times simulated.

\subsection{Higher correlation functions: charge-charge correlators}

\begin{figure}
    \centering
	\includegraphics[scale = 0.87]{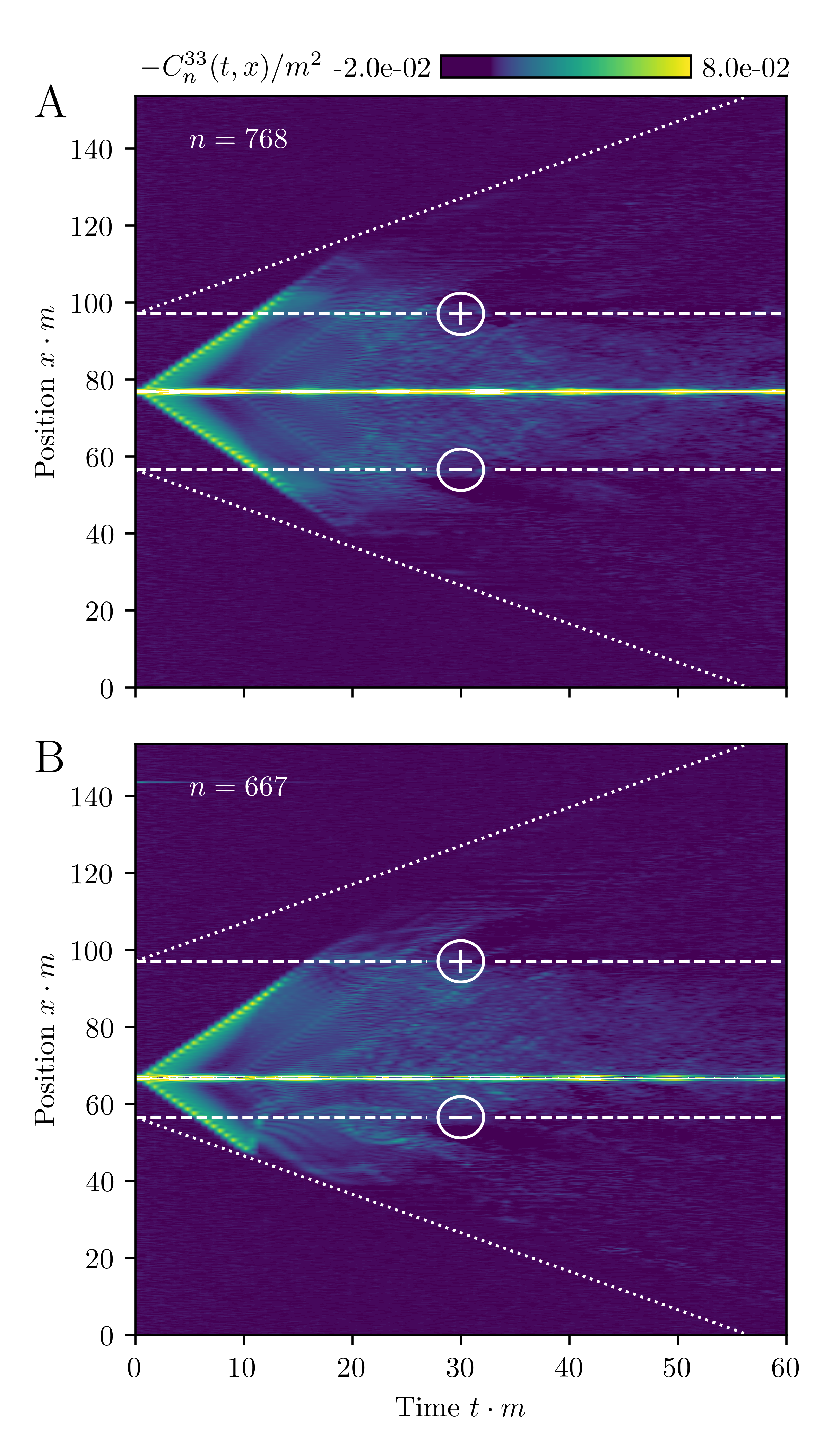}
	\caption{Connected charge-charge correlation function $C_n^{33}(t,x)$. \textbf{Panel A}: Reference site $n=768$. \textbf{Panel B}: Reference site $n=667$. White data points correspond to clipped data outside the given color range. Dashed lines denote positions of the external charges; dotted lines denote the outer boundaries of initial future light cones of the external charges. Parameters read $N_1=1536$, $L=153.6/m$, $a=0.1/m$, $a_t = 0.02\, a$, $g/m=1$, NLO improvements, string length $d=40.5/m$, external charges $N_0^3 = 8.0$, 3 sampling runs averaged.}\label{Figj0j0}
\end{figure}

In this subsection we study charge-charge correlations involving four fermion fields. This is used to further analyze the color-charge accumulations in the interior of supercritical strings, whose existence is inferred in the previous section from the fermion charge distribution and gauge-invariant fermion numbers of Fig.~\ref{FigStrings}, involving expectation values of two fermion fields. The quantity of interest is the connected anticommutator 
\begin{equation}
C_{nm}^{ab}(t) = \frac{1}{2} \langle \{\rho_n^a (t), \rho_m^b(t)\} \rangle - \langle \rho_n^a(t)\rangle \langle \rho_m^b(t)\rangle,
\label{eq:rr}
\end{equation}
which is computed using Eq.~(\ref{ConnectedChargeChargeCorrelFunctionModes}). This quantity is not gauge invariant, but turns out to be very suitable for discussing some characteristic nonequilibrium phenomena in QCD$_{1+1}$ associated to four-fields correlations. For all results shown, initial vacuum fluctuations have been taken into account.

In the following we consider $C_{nm}^{ab}(t) \equiv C_{n}^{ab}(t,x)|_{x \equiv m a}$ as a function of time $t$ and position $x$ for given reference site $n$ or position $x = n/(10m)$ for the parameters employed.
In Fig.~\ref{Figj0j0}, $C_{n}^{33}(t,x)$-data are displayed both for reference site $n=768$ and reference site $n=667$. The external charges are $N_0^3 = 8.0$ at position $x_-\simeq 57/m$ and at $x_+\simeq 97/m$, respectively, such that both reference sites are located between the charges. 

One observes that charge-charge correlations spread in space as time is increasing, propagating beyond the external charges. The correlations are sharply peaked at equal points $x/a=n$. In a first temporal regime correlations spread with approximately twice the speed of light and less along the string and beyond, with a comparably large front peak value. The regime of propagation with twice the speed of light lasts until correlations have reached part of the boundary of the external charges' future light cones that lies outside the string. Subsequently, a second temporal regime begins, in which correlations propagate with a maximum front velocity of approximately the speed of light. 

Since the correlation (\ref{eq:rr}) represents an anticommutator (as opposed to commutator) expectation value of two bosonic composites, the propagation can exceed the speed of light without being in conflict with any fundamental principle. We have seen above in Fig.~\ref{FigStrings} that the color charges can move close to the speed of light and in opposite directions. If the charge-charge anticommutator expectation value is approximately a function of the spatial difference, then the spreading of correlations within the medium would exhibit approximately a maximum velocity of twice the speed of light as observed. In fact, once the light cone boundaries of the external charges are reached, since there is no medium outside the light cones, there is no relative motion possible between the fermions inside and outside the light cones and the maximum correlation spreading drops down to the speed of light. We emphasize that this observation is a genuine nonequilibrium phenomenon, which would not be possible in thermal equilibrium where equal-time correlators are constant and anticommutator and commutator expectation values are related by the fluctuation-dissipation relation~\cite{Berges:2004yj}.

\begin{figure}
    \centering
	\includegraphics[scale = 0.87]{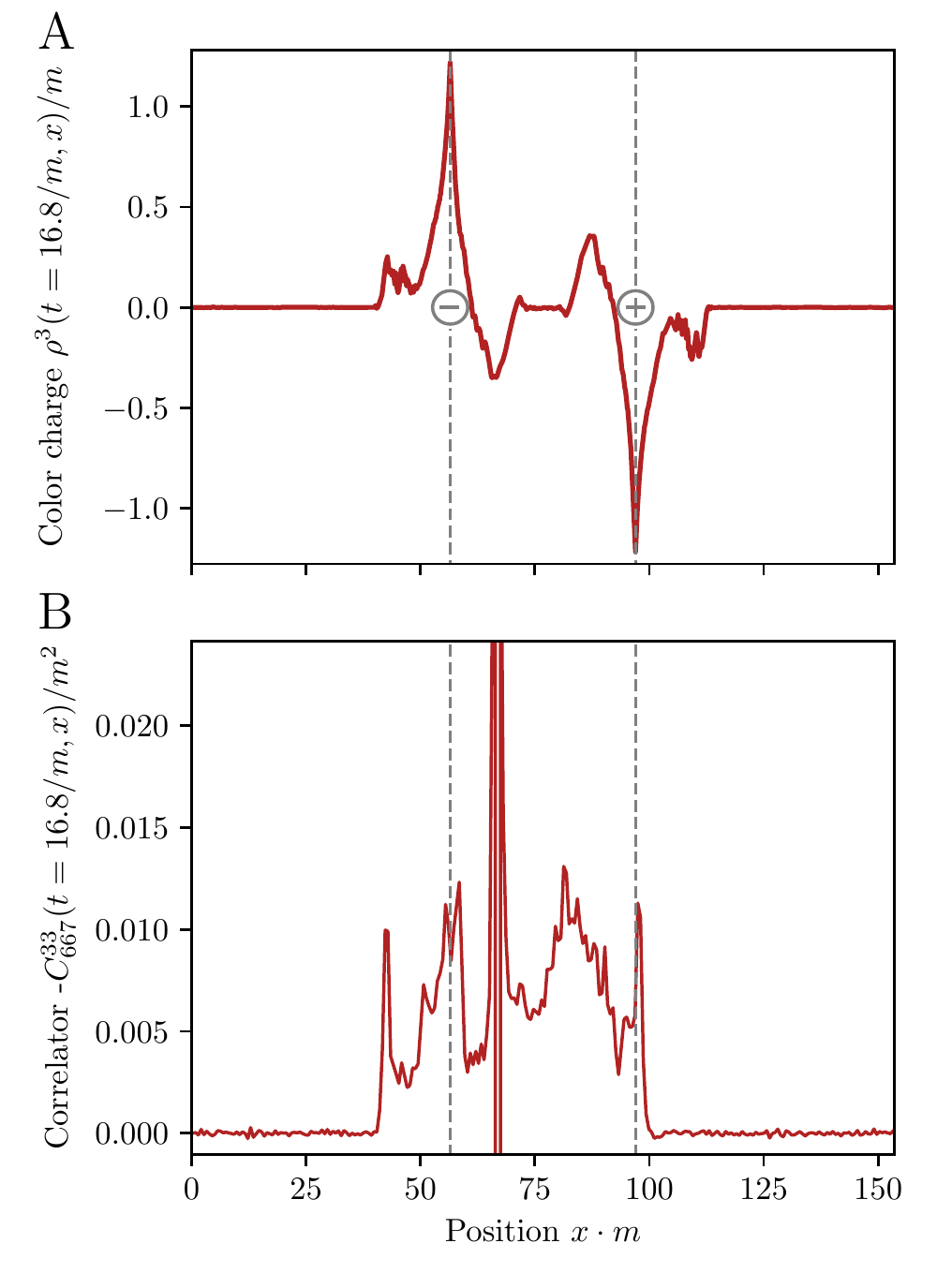}
	\caption{\textbf{Panel A}: Color charge $\rho^3(x)$. \textbf{Panel B}: Connected charge-charge correlator $C_{n}^{33}(t=18/m,x)$ at reference site $n=667$; $C_{n}^{33}$-data is averaged among each six lattice sites to smooth occurring oscillations among neighboring lattice sites. Both plots show data at time $t = 16.8/m$. Dashed lines denote positions of the external color charges. Parameters read $N_1=1536$, $L=153.6/m$, $a=0.1/m$, $a_t = 0.02\, a$, $g/m=1$, NLO improvements, string length $d=40.5/m$, external charges $N_0^3 = 8.0$, 3 sampling runs averaged.}\label{rho_time_18}
\end{figure}

Having globally studied connected $\rho\rho$-correlations, we now take a closer look at an example Cauchy surface of constant time with the aim of relating the correlation peaks and falls to the color charge accumulations appearing. In Fig.~\ref{rho_time_18}A the color charge 3-component is displayed at time $t=16.8/m$. Apart from the peaks around the external color charges at the denoted positions $x_- \equiv x_2\simeq 57/m$ and $x_+ \equiv x_5\simeq 97/m$, four dominant accumulations are present at positions $x_1 \simeq 43/m, x_3\simeq 68/m, x_4\simeq 86/m, x_6\simeq 111/m$. A simple picture of the external color charges being gradually screened could only explain the peaks around the external charges, not the four additional ones.

In Fig.~\ref{rho_time_18}B the connected $\rho\rho$-correlation function $C_{n}^{33}(t,x)$ is displayed at time $t=16.8/m$ and reference site $n=667$. For illustration purposes its negative values are shown. Primarily, we notice that the correlation function peaks match the color charge accumulations. Since the reference site $n=667$ is lying inside the color charge accumulation around position $x_3$, the pictured correlation function $C_{667}^{33}(t,x)$ can be interpreted to measure the charge-charge correlation at position $x$ with the accumulation around position $x_3$ at time $t=16.8/m$. We clearly find that the peak at $x_3$ is correlated with the color charge accumulations at positions $x_1,x_2,\dots, x_5$, but not with the accumulation around $x_6$ at time $t=16.8/m$ yet. The reason for this can be seen in Fig.~\ref{Figj0j0}B: At the time displayed $\rho\rho$-correlations did not propagate yet to the vicinity of $x_6$. Indeed, at the respective later times we find also a nonvanishing value for the correlation function at $x_6$, such that the color charge accumulations arising inside and around a supercritical color string are correlated with each other. As the data in Sec.~\ref{SectionStringBreaking} already suggest, these multiparticle excitations become increasingly fragmentated with time increasing.

\section{Conclusions}\label{SectionConclusions}
In the present study we investigated real-time fermion production via the non-Abelian Schwinger mechanism as well as the dynamical breaking of color strings between external static color charges in QCD$_{1+1}$ with gauge group $SU(2)$ using classical-statistical reweighting techniques. Within this setting we confirmed available analytic results and demonstrated that lattice improvements up to second order can significantly improve convergence towards the continuum limit of nonvanishing fermion numbers in homogeneous configurations. We observed non-Abelian plasma oscillations upon including the backaction of created fermions onto the gauge sector. For $SU(2)$ gauge theory plasma oscillations half of the produced fermions and antifermions is accelerated in positive spatial direction, the other half simultaneously in negative spatial direction. By means of improving the fermion current which feeds back to the chromoelectric field, the fermion number convergence behavior in non-Abelian plasma oscillations at late times can be optimized using higher-order lattice improvement terms, as well.

Additionally, we studied inhomogeneous initial conditions, focusing on configurations with a color string stretched between two external color charges. Being far from equilibrium, non-Abelian string breaking gives rise to a wide variety of involved dynamical processes, including fermion pair production, charge screening, plasma oscillations, fermions and antifermions continuously occupying the surrounding space around the string --- all of which emerging from apparently simple initial conditions. In particular, we showed that within supercritical color strings dynamical color charge accumulations can arise. Employing a connected charge-charge correlation function we demonstrated that these charge accumulations are correlated, with correlations propagating initially with almost twice the speed of light. 

These phenomena may become accessible in quantum simulators in the not too distant future, if efficient implementations become available. Improved lattice Hamiltonian formulations as those employed in this work can be a crucial ingredient, since significantly smaller lattices may be employed to obtain physical results. This becomes particularly important going beyond the one-dimensional benchmark case, when the real-time confinement dynamics is no longer of geometric origin but arises solely from the non-Abelian character of the gauge theory.   

\begin{acknowledgments}
We thank P.~Hauke, V.~Kasper, N.~Müller, R.~Ott, J.~Pawlowski, J.~Schneider, and T.~Zache for discussions and collaborations on related work. D.~Spitz receives support from the Konrad-Adenauer Foundation. This work is part of and supported by the DFG Collaborative Research Centre “SFB 1225 (ISOQUANT)”.
\end{acknowledgments}

\appendix

\section{Aspects of the lattice setup}

\subsection{Fermion initial conditions}\label{AppendixFermionicInitialConditions}
Initially, we implement free fermions with $n^u_{q,a} (0) = n^v_{q,a} (0) = 0$ for all $q,a$, using the free solution to the Dirac equation:
\begin{subequations}
\begin{align}
\phi_{n,q,a}^{u, \alpha} (0) &= u_{q}^\alpha \exp \bigg(+\frac{2\pi i \tilde{q} n}{N_1}\bigg)\,,\\
\phi_{n,q,a}^{v, \alpha} (0) &= v_{q}^\alpha \exp \bigg(-\frac{2\pi i \tilde{q} n}{N_1}\bigg)\,,
\end{align}
\end{subequations}
the $u$- and $v$-eigenspinors reading
\begin{subequations}
\begin{align}
u_q^\alpha &= \frac{1}{\sqrt{2\omega_q }}\left(\begin{matrix}
+\sqrt{\omega_q + p_q}\\
+\sqrt{\omega_q - p_q}
\end{matrix}\right),\\
v_q^\alpha &= \frac{1}{\sqrt{2\omega_q }} \left(\begin{matrix}
+\sqrt{\omega_q + p_q}\\
-\sqrt{\omega_q - p_q}
\end{matrix}\right).
\end{align}
\end{subequations}
$\omega_q$ is computed from $m_q,\,p_q$ using Eqs. (\ref{FreePhysicalMass}) and (\ref{FreePhysicalMomentum}) via 
\begin{equation}
\omega_q^2 = m_q^2+p_q^2\,.
\end{equation}

\subsection{Bosonic quantum fluctuations}\label{AppendixBosonicQuantumFluctuations}
Here we describe the initial bosonic quantum fluctuations, added to the classical gauge string configuration investigated in Sec.~\ref{SectionInhomogeneous}.

We denote creation and annihilation operators of a gauge boson with lattice momentum $\tilde{q}=-N_1/2,\dots,N_1/2-1$ and color index $a$ by $a_q^{a,\dagger},a_q^{\mathstrut a}$, respectively. The quantum fluctuations to construct correspond to the quantum-half contribution to bosonic vacuum occupation numbers, thus requiring
\begin{subequations}
\begin{align}
\langle a^{\mathstrut a}_q \rangle &= \langle a^{a,\dagger}_q\rangle = 0,\\
\frac{1}{2}\langle \{ a^{a,\dagger}_q,a^{\mathstrut a}_q\}\rangle &= \frac{1}{2}\Theta (Q-2\pi\tilde{q}/L)\Theta (Q+2\pi\tilde{q}/L).
\end{align}
\end{subequations}
In order to stay well below the lattice UV cutoff, we introduced a finite momentum scale $Q$ up to which fluctuations are taken into account, merely. We carefully checked for insensitivity of the obtained results to the precise choice of $Q$ and consistently specified $Q=\pi/(3a)$.
The quantum dynamics of $a_q^{a,\dagger},a_q^{\mathstrut a}$ are in the classical-statistical approximation imposed by sampling complex numbers $\alpha_q^{a,*},\alpha_q^{\mathstrut a}$ from a given distribution function and computing expectation values of $a_q^{a,\dagger},a_q^{\mathstrut a}$ according to the latter. As a distribution function $\mathcal{W}(\alpha_q^{\mathstrut a},\alpha_q^{a,\dagger})$ we chose the standard Gaussian distribution, $\bar{\alpha}_q^a$ denoting its average value and $\sigma^a_q$ its width,
\begin{equation}
\mathcal{W}\big(\alpha_q^{\mathstrut a},\alpha_q^{a,\dagger}\big) = \frac{1}{2\pi (\sigma^a_q)^2}\exp \Bigg( -\frac{|\alpha_q^a - \bar{\alpha}_q^a|^2}{2(\sigma^a_q)^2} \Bigg).
\end{equation}
The imposed one- and two-point functions for quantum-half fluctuations then read
\begin{subequations}
\begin{align}
\langle a^{\mathstrut a}_q \rangle &= \int \textrm{d}\alpha^{\mathstrut a}_q\, \textrm{d}\alpha^{a,*}_q\; \mathcal{W}\big(\alpha_q^{\mathstrut a},\alpha_q^{a,\dagger}\big) \alpha_q^{\mathstrut a} \nonumber\\
& = 0,\\
\frac{1}{2}\langle \{ a^{a,\dagger}_q,a^{\mathstrut a}_q\}\rangle &= \int \textrm{d}\alpha^{\mathstrut a}_q\, \textrm{d}\alpha^{a,*}_q\; \mathcal{W}\big(\alpha_q^{\mathstrut a},\alpha_q^{a,\dagger}\big) \alpha_q^{a,*}\alpha_q^{\mathstrut a} \nonumber \\
&= \frac{1}{2} \Theta (Q-2\pi \tilde{q}/L)\Theta (Q+2\pi \tilde{q}/L),\
\end{align}
\end{subequations}
resulting in $\bar{\alpha}_q^a = 0$ and $\sigma^a_q = \Theta (Q-2\pi\tilde{q}/L)\Theta(Q+2\pi\tilde{q}/L)/\sqrt{2}$. $\alpha^{a,*}_q$ and $\alpha^{\mathstrut a}_q$ being drawn randomly according to $\mathcal{W}(\alpha_q^{\mathstrut a},\alpha_q^{a,\dagger})$, we compute their Fourier-transformed spatial counterparts as
\begin{equation}
\alpha_n^a = \frac{1}{L}\sum_{q\in \tilde{\Lambda}}\alpha_q^a e^{ 2\pi i \tilde{q}n/N_1}.
\end{equation}
Fluctuations of the gauge potential and the chromoelectric field, added to the classical initial conditions, are computed from this as
\begin{equation}
\delta A_n^a = \sqrt{2a^2}\;\textrm{Re}(\alpha_n^a),\qquad \delta E_n^a = \sqrt{2} \; \textrm{Im} (\alpha^a_n).
\end{equation}
Since link variables are classically initialized to unity in this work, upon inclusion of classical-statistical sampling they initially read
\begin{equation}
U_n(t=0) = \exp \big( iga \, \delta A_n^a \, T^a\big).
\end{equation}
Observables are computed at each time step as ensemble averages of a given number of simulated runs with different fluctuating initial conditions generated this way.

\subsection{Deriving equations of motion}\label{AppendixEOMS}
Crux of deriving equations of motion for the involved field degrees of freedom are commutation and anticommutation relations between the fields to finally employ $i\partial_t O = [O,H]$ for a field operator $O$.

The chromoelectric field equation of motion follows from using
\begin{subequations}
\begin{align}\label{CommutationRelationsBosons}
\big[E_n^a, A_m^b\big] &= \frac{i}{a} \,\delta^{\mathstrut}_{n,m} \delta^{\mathstrut}_{a,b}\,,\\
\big[A_n^a, A_m^b\big] &= \big[E_n^a,E_m^b\big] = 0\,,
\end{align}
\end{subequations}
in order to obtain by application of Baker-Campbell-Hausdorff
\begin{subequations}\label{CommutationRelationsEU}
\begin{align}
\big[E_n^a,U^{\mathstrut}_m\big] &= -g \,\delta^{\mathstrut}_{n,m}\,T^a \,U^{\mathstrut}_{n} \,,\label{CommutationRelationsEU:a}\\
\big[E_n^a,U^\dagger_m\big]      &= +g \, \delta^{\mathstrut}_{n,m} \, U^\dagger_{n}\, T^a\,.\label{CommutationRelationsEU:b}
\end{align}
\end{subequations}
Clearly, we find $[E_n^a, H^{(\text{g})}]=0$ for all $n,a$. It remains to compute $[E_n^a, H^{(\text{f})}]$, for which we make extensive use of relations (\ref{CommutationRelationsEU:a}) and (\ref{CommutationRelationsEU:b}). Expressions such as
\begin{align}
\big[E_n^a, U_{m} &U_{m+1} U_{m+2}\big] \nonumber\\
=\,& -g\,\big(\delta_{n,m} \,T^a\, U_{m}U_{m+1}U_{m+2}\nonumber\\
& + \delta_{n,m+1} \,U_{m}\,T^a\, U_{m+1}U_{m+2} \nonumber\\
& + \delta_{n,m+2}\,U_{m}U_{m+1}\,T^a\, U_{m+2}\big)
\end{align}
occur.

The link variable equation of motion, Eq. (\ref{EOMLinkVariables}) follows easily from $[U_n,H^{(\text{g})}]$ and
\begin{equation}
\big[U_n,E_m^aE_m^a\big] = \big[U_n,E_m^a\big] \,E_m^a + E_m^a \,\big[U_n, E_m^a\big].
\end{equation}

To obtain equations of motion for the fermion mode functions we evaluate $[\psi_n,\,H^{(\text{f})}]$. We make use of the algebraic identity
\begin{equation}
\big[\psi_{n,c}^\gamma ,\,\psi_{l,a}^{\dagger,\alpha}\psi_{m,b}^{\dagger,\beta}\big] = \big\{\psi_{n,c}^\gamma, \psi_{l,a}^{\dagger,\alpha}\big\}\,\psi_{m,b}^{\dagger,\beta} - \psi_{l,a}^{\dagger,\alpha}\,\big\{\psi_{n,c}^\gamma, \psi_{m,b}^{\dagger,\beta}\big\}\,,
\end{equation}
where Dirac and color indices have explicitly been restored. Using this identity, straightforwardly the equation of motion (\ref{EOMFermionModes}) follows.

\subsection{Volume dependence of substructures in plasma oscillations}\label{AppendixInfraredArtefacts}

\begin{figure}
    \centering
	\includegraphics[scale = 0.87]{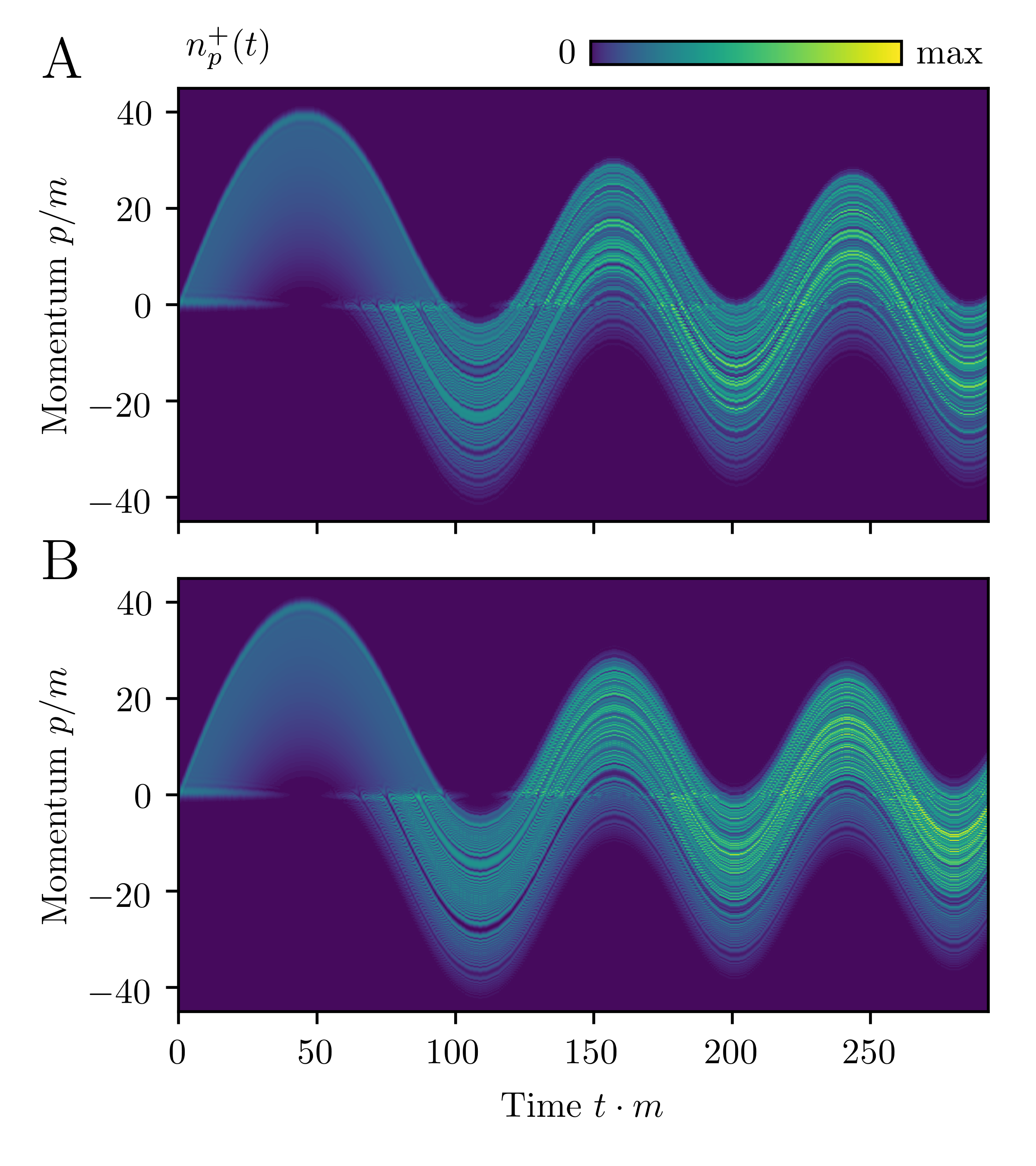}
	\caption{Time-evolution of the Abelianized fermion number momenta spectrum $n^{+}_p(t)$. \textbf{Panel A}: $N_1=512$. \textbf{Panel B}: $N_1=768$. The employed lattice parameters of both panels read $L=38.4/m$, $a=0.05/m$, $a_t=0.02 \,a$, $g/m=0.3$, NNLO improvements, with an initial chromoelectric field $\epsilon = (3,0,0)$.}\label{FigInfraredLatticeArtefacts}
\end{figure}

In Sec. \ref{SecPlasmaOscillations} we observed that substructures occur in momentum space-resolved fermion numbers as a result of temporary fermion production dropouts in the zero-momentum mode. In Fig. \ref{FigInfraredLatticeArtefacts} Abelianized fermion numbers $n^{+}_p(t)$ are displayed again, now for both $N_1=512$ and $N_1=768$. We primarily notice that structures are more present at smaller lattice size. Related to finite volume and being smaller at larger lattice size, we take this as a hint that the occurring fermion production dropouts are, essentially, infrared lattice artifacts, disappearing continuously with increasing lattice sizes.

\subsection{Comparing Abelianized and gauge-invariant fermion numbers}\label{AppendixCompareFermionNumbersDefs}

\begin{figure}
    \centering
	\includegraphics[scale = 0.87]{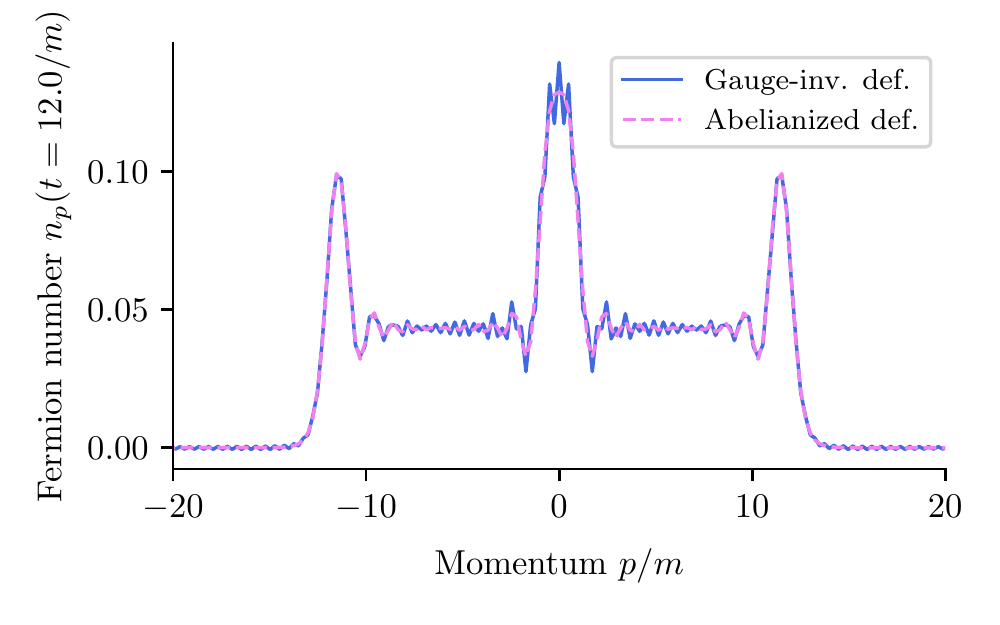}
	\caption{Abelianized fermion number momentum spectrum, both diagonalized colors summed (violet, dashed line), compared to the gauge-invariant fermion number momentum spectrum (blue line) at time $t = 12/m$. The employed lattice parameters of both panels read $L=25.6/m$, $a=0.05/m$, $a_t=0.02 \,a$, $g/m=0.1$, NNLO improvements, constant chromoelectric field $\epsilon = (2,0,0)$.}\label{GaugeInvVsDiagFermionNumbers}
\end{figure}

A proof of consistency within our work is to compare both the Abelianized and the gauge-invariant fermion number definitions. In Fig. \ref{GaugeInvVsDiagFermionNumbers} we find the two fermion number definitions roughly match each other, comparing fermion number momentum spectra for the two in a constant chromoelectric background field. Merely, the gauge-invariant fermion numbers oscillate around the Abelianized ones with amplitudes varying in time. At times, oscillations are larger than in Fig. \ref{GaugeInvVsDiagFermionNumbers}, at times smaller, even nearly vanishing.

Enabling backcoupling of the created fermions onto the gauge sector, we find that both fermion number definitions show the same oscillating pattern. Fermion number momentum spectra and total fermion numbers computed from both definitions agree with each other, though not displayed here.

Furthermore, comparing to the gauge-invariant fermion numbers in Hebenstreit \emph{et al.} \cite{Hebenstreit:2013qxa}, we note that fluctuations in the momentum spectra computed from our simulations are sometimes much larger than those encountered for example in Fig. 3 of \cite{Hebenstreit:2013qxa}. We believe the reason for this to be the different type of lattice fermions employed: While we use Wilson fermions in the present work, Hebenstreit \emph{et al.} implemented low-cost fermions. We expect that if in our Wilsonian approach we employed classical-statistical sampling also in the homogeneous setting around a coherent initial chromoelectric field, with increasing the number of samples fluctuations in gauge-invariant fermion number momentum spectra would continuously vanish.

\section{Analytic $SU(2)$ pair production results}\label{AppendixAnalytics}
Besides the original computation by Schwinger using the one-loop effective action \cite{Schwinger1951}, it is possible to solve the Dirac equation in homogeneous, constant background $U(1)$ gauge fields using quantum kinetic theory \cite{Hebenstreit:2010vz}. Analogously, we can proceed in $SU(2)$ gauge theory upon diagonalizing color degrees of freedom \cite{Nayak:2005pf}. Most of the derivation proceeds similarly to \cite{Gelis:2015kya, Hebenstreit:2010vz}, reducing the number of space-time dimensions to 2. We solely sketch differences here.

We define an inner product for arbitrary mode functions $\phi$, $\chi$ as
\begin{equation}
(\phi_a|\chi_b) \equiv \int dx \; \phi^\dagger_a (t,x)\,\chi_b^{\mathstrut} (t,x)\,,
\end{equation}
with color indices $a,b$ restored. In what follows ``in'' specifies a free fermionic initial state at time $t\to -\infty$; ``out'' labels the asymptotically free but gauge-rotated fermionic final state at time $t\to \infty$. At $t\to -\infty$ we expand the Dirac field color components $\psi_a$ in momentum modes,
\begin{equation}
\psi_a (x) = \int \frac{dp}{2\pi}\, \bigg(\phi^{\text{in},+}_{p,a} (x)\: b^{\text{in}}_{p,a} + \psi^{\text{in},-}_{p,a} (x)\: d^{\text{in},\dagger}_{p,a} \bigg)\,.
\end{equation}
We expand $\psi_a$ as well in out-state momentum modes,
\begin{equation}
\psi_a (x) = \int \frac{dp}{2\pi}\, \bigg(\phi^{\text{out},+}_{p,a} (x)\: b^{\text{out}}_{p,a} + \psi^{\text{out},-}_{p,a} (x)\: d^{\text{out},\dagger}_{p,a} \bigg)\,,
\end{equation}
and find the Bogoliubov transformation coefficient
\begin{equation}
\beta(p,a) \equiv \frac{1}{2\omega_p}\left(\phi_{p,a}^{\text{out},+}\left|\phi_{-p,a}^{\text{in},-}\right)\right.,
\end{equation}
in order to arrive with $\beta(p)\equiv (\beta(p,1),\beta(p,2))$ at
\begin{eqnarray}
\mathcal{F}(p)\equiv \mathcal{F}(t\rightarrow \infty,p) &\equiv& \lim_{V\to \infty} \frac{1}{V}\sum_a  \big\langle 0_{\text{in}}\big|a_{p,a}^{\text{out},\dagger}a_{p,a}^{\text{out}}\big|0_{\text{in}}\big\rangle \nonumber\\
&=& \lim_{V\to \infty}\text{Tr}\big(\beta^\dagger(p)\,\beta(p)\big),
\end{eqnarray}
for the in-vacuum expectation value of the out-particle number density operator. Here, $V$ denotes configuration space volume, finally taken to infinity. We note that if it is possible to Abelianize color degrees of freedom, $\mathcal{F}(p)$ is left invariant upon the diagonalization scheme.

Restricting to a uniform, constant chromoelectric field $E^a_n(t) = En^a$ with $n^a n^a =1$ for all $n\in\Lambda$, we can diagonalize color degrees of freedom as in Sec. \ref{SubsectionAbelianization}. We find a diagonalized chromoelectric field $\text{diag}(\epsilon/2,-\epsilon/2)$ in units of $E_c$ and again label diagonalized color components by $\pm$, corresponding to effective electric fields $\pm E/2$. Having applied the diagonalization procedure, we proceed as in $U(1)$ gauge theory \cite{Hebenstreit:2010vz} to finally find
\begin{widetext}
\begin{eqnarray}\label{EqAppendixAnalyticsFp}
\mathcal{F}(p)	&=&	\sum_{\pm}\frac{1}{8}\left(1+\frac{\hat{p}^\pm}{\sqrt{2\eta^\pm+(\hat{p}^\pm)^{2}}}\right)e^{-\pi\eta^\pm/4}\left|\left(\sqrt{2\eta^\pm+(\hat{p}^\pm)^{2}}-\hat{p}^\pm\right)D_{-1+i\eta^\pm/2}\left(-\hat{p}^\pm\,e^{-i\pi/4}\right)\right.\nonumber\\
& & \qquad \qquad \qquad \qquad \qquad \qquad \qquad \qquad \quad \left.-2e^{i\pi/4}D_{i\eta^\pm/2}\left(-\hat{p}^\pm\,e^{-i\pi/4}\right)\right|^{2},
\end{eqnarray}
\end{widetext}
with
\begin{subequations}
\begin{equation}
\hat{p}^\pm \equiv \sqrt{\pm\frac{4}{\epsilon}} \frac{p\pm gE t/2}{m}\,,
\end{equation}
and
\begin{equation}
\eta^\pm \equiv \pm \frac{2}{\epsilon}\,.
\end{equation}
\end{subequations}
We denote the individual addends $\pm$ in Eq. (\ref{EqAppendixAnalyticsFp}) by $\mathcal{F}^\pm(p)$. $\mathcal{F}^\pm(p)$ approaches a nonvanishing constant value for large $p$,
\begin{equation}
\lim_{p\to \infty} \mathcal{F}^\pm (p) = \exp \bigg(-\frac{2\pi}{\epsilon}\bigg),
\end{equation}
which translates, approximately, into a constant rate per spatial volume $L$ and per Abelianized color direction $\pm$ at which fermion-antifermion pairs are created \cite{Cohen:2008wz, Tanji:2008ku},
\begin{eqnarray}
\frac{\dot{n}^\pm}{L} &=& \frac{1}{T}\int \frac{dp'}{2\pi} \; \Theta (p') \; \Theta(gET/2-p')\lim_{p\to\infty} \mathcal{F}^{\pm}(p) \nonumber \\
&=& \frac{m^2\epsilon}{4\pi} \exp\bigg(-\frac{2\pi}{\epsilon}\bigg).
\end{eqnarray}
The physically measurable, total production rate of fermions and antifermions per volume $L$ is thus given by
\begin{equation}
\frac{\dot{n}}{L} \equiv \frac{\dot{n}^+ + \dot{n}^-}{L} = \frac{m^2\epsilon}{2\pi}\exp\bigg(-\frac{2\pi}{\epsilon}\bigg).
\end{equation}

\bibliography{literature}

\end{document}